\documentclass[aps,pra,onecolumn,reprint,floatfix, superscriptaddress]{revtex4-2}

\usepackage{subfigure}
\usepackage{psfrag,graphicx}
\usepackage{pict2e}
\usepackage{dcolumn}
\usepackage{amsmath,amssymb}
\usepackage{bm}
\usepackage{xcolor}
\usepackage{latexsym}
\usepackage{epstopdf}
\usepackage[english]{babel}
\UseRawInputEncoding
\usepackage{amsfonts}
\usepackage{bm}
\usepackage{natbib}
\usepackage{bbold}

\usepackage{enumerate}
\definecolor{myblue}{rgb}{0.2,0.2,0.8}
\definecolor{myzard}{cmyk}{0,0,0.05,0}
\definecolor{mywhite}{rgb}{1,1,1}
\definecolor{myred}{rgb}{1,0.,0.3}
\usepackage[colorlinks=true,citecolor=myblue,linkcolor=myred]{hyperref}

\definecolor{mygrey}{gray}{0.35}
\definecolor{myblue}{rgb}{0.2,0.2,0.8}
\definecolor{myzard}{cmyk}{0,0,0.05,0}
\definecolor{mywhite}{rgb}{1,1,1}
\definecolor{mywhite}{rgb}{1,1,1}
\definecolor{myred}{rgb}{1,0.,0.3}

\def\be{\begin{equation}}
\def\ee{\end{equation}}
\def\ba{\begin{align}}
\def\enda{\end{align}}
\def\bi{\begin{itemize}}
\def\ei{\end{itemize}}

\def\beq{\begin{equation}}
\def\beq{\begin{equation}}
\def\eeq{\end{equation}}

 \newcommand{\ket}[1]{|#1\rangle}

\newcommand\Tr{\mathrm{Tr}}
\newcommand{\mean}[1]{\langle #1\rangle}

\begin{document}

\title{Decimation technique for open quantum systems: a case study with driven-dissipative bosonic chains}

\author{\'{A}lvaro G\'{o}mez-Le\'{o}n}
\email{a.gomez.leon@csic.es}
\affiliation{Institute of Fundamental Physics IFF-CSIC, Calle Serrano 113b, 28006 Madrid, Spain.}

\author{Tom\'{a}s Ramos}
\affiliation{Institute of Fundamental Physics IFF-CSIC, Calle Serrano 113b, 28006 Madrid, Spain.}

\author{Diego Porras}
\affiliation{Institute of Fundamental Physics IFF-CSIC, Calle Serrano 113b, 28006 Madrid, Spain.}

\author{Alejandro~Gonz\'{a}lez-Tudela}
\email{a.gonzalez.tudela@csic.es}
\affiliation{Institute of Fundamental Physics IFF-CSIC, Calle Serrano 113b, 28006 Madrid, Spain.}

\begin{abstract}
The unavoidable coupling of quantum systems to external degrees of freedom leads to dissipative (non-unitary) dynamics, which can be radically different from closed-system scenarios. Such open quantum system dynamics is generally described by Lindblad master equations, whose dynamical and steady-state properties are challenging to obtain, especially in the many-particle regime. Here, we introduce a method to deal with these systems based on the calculation of (dissipative) lattice Green's function with a real-space decimation technique. Compared to other methods, such technique enables obtaining compact analytical expressions for the dynamics and steady-state properties, such as asymptotic decays or correlation lengths. We illustrate the power of this method with several examples of driven-dissipative bosonic chains of increasing complexity, including the Hatano-Nelson model. The latter is especially illustrative because its surface and bulk dissipative behavior are linked due to its non-trivial topology, which manifests in directional amplification.
\end{abstract}
\maketitle

\section{Introduction}

Closed quantum systems are governed by a Hermitian evolution dictated by the Hamiltonian. These systems, however, generally correspond to idealizations used to study more complex scenarios. 
In practical situations, perfect isolation is not possible and quantum systems interact with external degrees of freedom. 
This interaction leads to non-unitary (dissipative) dynamics~\cite{breuer-petruccione}, very different from the closed-system case. 
Understanding such dissipative dynamics is of utmost importance, e.g., for  quantum technologies, because it establishes bounds on their performance~\cite{preskill18a}, and, if properly engineered, can even be used as resource for them~\cite{diehl08a,verstraete09a,Paulisch2016,Ramos14a,Asenjo-Garcia2017a}. 

The description of open quantum systems needs to be upgraded from a wave-function formalism to a density matrix one, whose dynamics is well captured by effective Lindblad master equations~\cite{lindblad76a,Gorini2008}. Under the Born-Markov assumptions of weak system-bath couplings and memory-less baths, such master equations have a time-local form~\cite{breuer-petruccione}. However, even with this simplification, calculating the steady-state and dynamical features of such systems is still an outstanding challenge, since closed-systems methods are not directly applicable (see Ref.~\cite{Weimer2021} and references therein for an updated review on the subject). Although there are some analytical techniques based on third quantization~\cite{Prosen2008,Prosen2010}, resummation of perturbative series~\cite{Li2014b,Li2016}, weak symmetries~\cite{McDonald2021a}, or flow equations~\cite{Rosso2020},  open quantum systems are generally characterized numerically, e.g., via exact diagonalization or time Density-Matrix-Renormalization-Group methods~\cite{Vidal2003,white04a,Feiguin2005,feiguin13a,Volokitin2017,prior2010,Tamascelli2019,Daley2009,Daley2004}, which prevent in many cases a simple understanding of the phenomena.

\begin{figure}
    \centering
    \includegraphics[width=1\columnwidth]{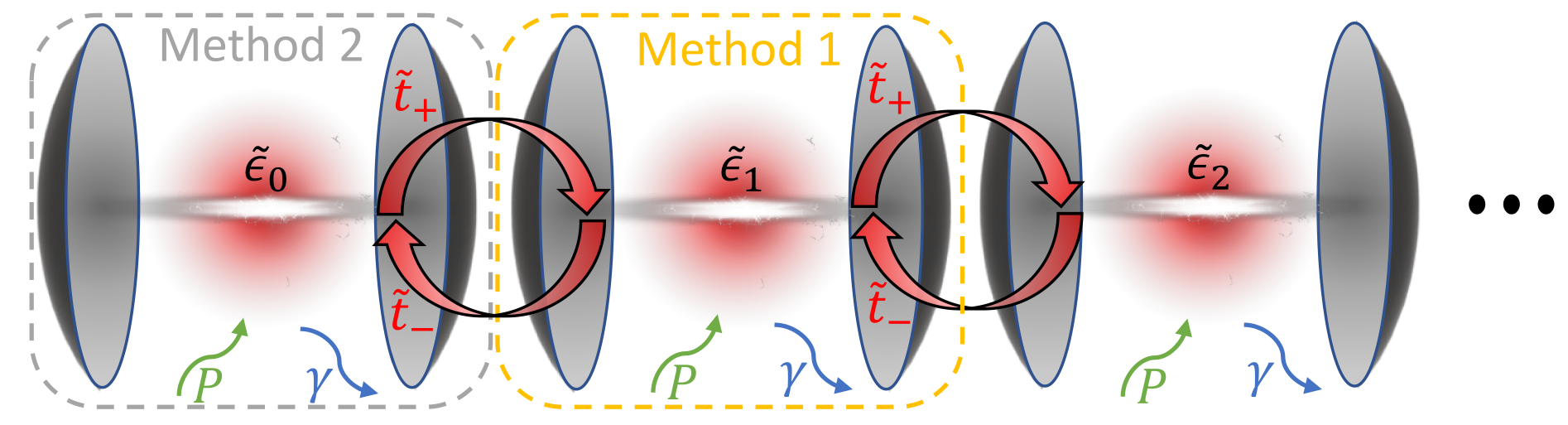}
    \caption{Schematic for the system: a bosonic chain with neighboring sites coupled by (coherent and incoherent) tunneling, $\tilde{t}_\pm$. Sites are also subject to local gain $P$ and loss $\gamma$ processes. The dashed rectangles represent the two different decimation schemes: method 1, based on decimating site $1$ until just site 0 remains; and method 2, based on adding sites at the edge until their Green's function remains constant with the number of sites. Method 1 is more appropriate to determine the Green's function for finite size systems, while method 2 is more effective in the semi-infinite limit.}
    \label{fig:Schematic}
\end{figure}

In this work, we introduce a method to characterize the dynamics and steady-steady properties of open quantum many-body systems based on the combination of (dissipative) lattice Green's functions~\cite{Economou2006} and a real-space decimation technique~\cite{Nelson1975,Odashima2016}, that can lead to compact analytical expressions in certain limits, i.e., semi-infinite chains. 

Lattice Green's functions were originally introduced in the condensed matter context to characterize the response functions of discrete systems in various dimensions and geometries~\cite{schwalm88a,schwalm1992,Schwalm1991,Dyson1953,Davison1970,Bass1985}. However, they have also been recently pointed out as a useful tool to characterize out-of-equilibrium situations subject to time-dependent drives~\cite{Arrachea2005,Arrachea2007} and/or coupling to environments, in combination with Keldysh formalism~\cite{Maghrebi2016,Sieberer2016,Gomez-Leon2021}. Decimation techniques~\cite{Nelson1975,Odashima2016} have already been applied to calculate lattice Green's functions in closed-system scenarios in one ~\cite{Ashraff1988,Chakrabarti1989,Southern1983,Liu1986} and higher-dimensional lattices~\cite{Southern1985,LopezSancho1985HighlyFunctions,Lewenkopf2013TheGraphene}, including some topological models~\cite{Ruocco2017,Peng2017BoundarySuperconductors,Pai2019}. 
Here, we show how to apply this technique in generic open quantum systems, where it has been scarcely used~\cite{Zhou2020}. The main advantage is that we will be able to find compact analytical expressions for the dissipative Green's function, without the need to calculate a matrix inversion. In particular, for the case of semi-infinite chains, we obtain that the two-point Green-function $G_{j,l}$ has two contributions: i) one that depends only on the relative distance $j-l$, and ii) one that depends on the distance from the edge, which encodes the surface effects.  We illustrate the power of this approach with several examples of increasing complexity of dissipative bosonic chains subject to gain and loss processes (see Fig.~\ref{fig:Schematic}), including one example which features topological quantum amplification, the Hatano-Nelson model~\cite{Hatano1996,Porras2019,Wanjura2020,Ramos2021}, where such separation between surface and bulk properties will be especially interesting.

The manuscript is structured as follows: in Section~\ref{sec:Green} we review the connection between (lattice) Green's functions and Lindblad master equations. In Section~\ref{sec:decimation}, we present two methods: the first one, discussed in~\ref{subsec:realspacedecimation}, calculates (dissipative) lattice Green's functions by direct decimation of all sites in real space; whereas the second method, introduced in~\ref{subsec:dyson}, takes advantage of the invariance of the system in the semi-infinite limit to solve the Dyson's equation. In addition, in~\ref{subsec:dynamics} we explain how to use this formalism to obtain the transient dynamics to the steady state. Then, in Section~\ref{sec:examples}, we illustrate this method with several examples of driven-dissipative bosonic chains, including physical ingredients until obtaining the Hatano-Nelson model, a paradigmatic model of topological amplification. Besides, we also show that this method can be combined with the input-output formalism~\cite{Ramos2021} to study the noise properties. 
Interestingly, with these tools we are able to obtain analytical expressions for its Green's function, its topological and noise properties, and, for the amplification dynamics, where we find features of two different dynamical regimes within the topological amplifying phase. Finally, in Section~\ref{sec:conclu}, we summarize our findings and point out to future work directions.

\section{Green's functions in open quantum systems~\label{sec:Green}}

Closed quantum system dynamics can be described by a wave-function, $\ket{\Psi(t)}$, whose evolution is governed by its Hamiltonian, $H$. 
In open quantum systems, the situation is more complex. First, the interaction with the environment requires upgrading from a wave-function to a density matrix description, which allows to describe mixed states. Besides, assuming that the dynamics of the bath can be adiabatically eliminated, under the Born-Markov conditions, the dynamics of the system's density matrix, $\rho$, is governed by the following time-local master equation~\cite{breuer-petruccione} (we take $\hbar=1$ in the rest of the manuscript):
\begin{equation}
\dot{\rho}=-i\left[H,\rho\right]+\sum_{\alpha}\gamma_{\alpha}\left[L_{\alpha}\rho L_{\alpha}^{\dagger}-\frac{1}{2}\left\{ L_{\alpha}^{\dagger}L_{\alpha},\rho\right\} \right],\label{eq:Lindblad11}
\end{equation}
which accounts for both the unitary evolution given by the system's Hamiltonian, $H$, and a non-unitary one induced by the Lindblad operators, $L_\alpha$, with associated rate $\gamma_\alpha$. Note, these Lindblad operators can describe both loss- and pump-type processes when $L_\alpha$ is proportional to annihilation and creation operators, respectively~\cite{Ramos2021}.

Obtaining the full density matrix operator $\rho(t)$ from Eq.~\eqref{eq:Lindblad11} is generally very challenging. However, this is generally not needed, because one is typically interested in certain experimental observables which admit simpler descriptions. For example, if one is interested in obtaining the mean values of certain operators, $\mean{O_j(t)}=\Tr\left[O_j\rho(t)\right]$, they can be shown to be given by:
\begin{equation}
\partial_{t}\mean{O_j}=i\mean{\left[H,O_j\right]}+\sum_{\alpha}\gamma_{\alpha}\left[\mean{ L_{\alpha}^{\dagger}O_j L_{\alpha}}-\frac{1}{2}\mean{ \left\{O_j,L_{\alpha}^{\dagger}L_{\alpha}\right\}}\right]\,.\label{eq:Lindblad2}
\end{equation}
Defining a vector with the mean values, $\mean{\mathbf{X}(t)}=(\mean{O_1(t)},\mean{O_2(t)},\dots)^T$, which includes all the operators coupled by Eq.~\eqref{eq:Lindblad2}, this equation can be written in matrix form as follows:
\begin{align}
\partial_t \mean{\mathbf{X}(t)}=-i\mathcal{D}\mean{\mathbf{X}(t)}\,,\,\label{eq:Lindblad3}
\end{align}
with $\mathcal{D}$ being the dynamical matrix, which is generally non-Hermitian and with complex eigenvalues. Note that in most cases the set of coupled differential equations of Eq.~\eqref{eq:Lindblad2} must be truncated to a finite number so that it can be numerically solved. 
Eq.~\eqref{eq:Lindblad3} can be formally solved as:
\begin{align}
\mean{\mathbf{X}(t)}=e^{-i\mathcal{D} t}\mean{\mathbf{X}(0)}\, ,\label{eq:Lindblad4}
\end{align}
with $\mean{\mathbf{X}(0)}$ the initial value of $\mean{\mathbf{X}}$.
Furthermore, assuming that the system is dynamically stable, that is, the eigenvalues of $\mathcal{D}$ have always a non-positive imaginary part, the steady state value of the operators can be obtained from $\mean{\mathbf{X}(t\to \infty)}$.

Other important quantities to characterize these systems are the so-called two-time averages $\mean{O_j(t+\tau)O^\dagger_l(t)}$, which are related to important experimental observables such as the spectrum or the correlation functions. Although they cannot be directly extracted from Eqs.~\eqref{eq:Lindblad2}-\eqref{eq:Lindblad3}, which are defined only for single-time averages, one can use the Quantum Regression Theorem~\cite{lax63a,breuer-petruccione} to show that they are governed by the same matrix $\mathcal{D}$, but with a different initial condition:
\begin{align}
\partial_\tau \mean{\mathbf{X}(t+\tau)O^\dagger_j(t)}=-i\mathcal{D} \mean{\mathbf{X}(t+\tau)O^\dagger_j(t)}\,.\,\label{eq:QRT1}
\end{align}
Besides, from a fundamental point of view, such double-time averages are also important because they are linked with the single particle Green's function of the system~\cite{Zubarev1960}, even in the open quantum system scenario~\cite{Sieberer2016}. 
Considering the standard definition for the bosonic, retarded Green's function~\cite{Economou2006}:
\begin{equation}
G_{j,l}\left(t+\tau,t\right)=-i\theta\left(\tau\right)\langle\left[\hat{O}_{j}\left(t+\tau\right),\hat{O}_{l}^{\dagger}\left(t\right)\right]\rangle\,,
\end{equation}
we find that Eq.~\eqref{eq:QRT1} translates to the following equation of motion for the Green's function:
\begin{equation}
    i\partial_{\tau}G(t+\tau,t)=\delta(\tau)+\mathcal{D} G(t+\tau,t)\ .\label{eq:GF-EOM1}
\end{equation}
Here, $G(t+\tau,t)$ is the Green's functions matrix with elements $G_{j,l}(t+\tau,t)$ and $\mathcal{D}$ is the dynamical matrix with the same coefficients as in the homogeneous differential equation, Eq.~\eqref{eq:Lindblad3}.
The steady state solution to Eq.~\eqref{eq:GF-EOM1} is obtained by means of a Fourier transform and a matrix inversion:
\begin{equation}
    G(\omega)=\left(\omega-\mathcal{D}\right)^{-1},\  \label{eq:resolvent}
\end{equation}
and is of great importance in dissipative quantum systems, because it can be used as resolvent for the equations of motion of various observables. Interestingly, the dynamical matrix $\mathcal{D}$ coincides with the effective Hamiltonian obtained from the Keldysh path integral method~\cite{Gomez-Leon2021}, and it can also be used for the characterization of topological properties.

In what follows, we describe how the decimation technique allows us to determine the dissipative Green's function without having to calculate the inverse of $\omega-\mathcal{D}$, which can be computationally advantageous for large systems. Furthermore, it will also allow us to find simple analytical expressions in certain cases, like in the semi-infinite limit, where it can be used to study bulk/boundary properties in topological models.
\section{Decimation for lattice Green's functions~\label{sec:decimation}}
For the sake of illustration, we will now particularize to the case of a one-dimensional chain, described by bosonic operators, $a_i^{(\dagger)}$, with $i=0,\dots, N-1$, satisfying $[a_i,a_j^\dagger]=\delta_{ij}$. Nevertheless, most of our results can be extrapolated to fermionic operators with small changes~\cite{Gomez-Leon2021}. 

We assume that the Hamiltonian describing the bosonic chain can be written as:
\begin{align}
    H=\sum_{j,l=0}^{N-1} t_{j,l}a_j^\dagger a_l\,,\label{eq:H}
\end{align}
where the Lindblad terms are given by both a (collective) decay term:
\begin{align}
\mathcal{L}_\mathrm{decay}[\rho]=\sum_{j,l}\gamma_{j,l}\mathcal{D}\left[a_j,a_l^{\dagger}\right]\,,\label{eq:decay}
\end{align}
and a pump term:
\begin{align}
\mathcal{L}_\mathrm{pump}[\rho]=\sum_{j,l}P_{j,l}\mathcal{D}\left[a_l^{\dagger},a_j\right]\,,\label{eq:pump}
\end{align}
with $\mathcal{D}[A,B](\rho)=A\rho B-\frac{1}{2}\left\{BA,\rho \right\}$. Note, that this allows us to capture very different models, such as (see Fig.~\ref{fig:Schematic}):
\begin{itemize}
    \item A simple dissipative coupled-cavity array with on-site energy $\epsilon$ and nearest-neighbor hopping $t_c$: $t_{i,j}=\epsilon \delta_{j,l} + t_c(\delta_{i,j+1}+\delta_{i,j-1})$, also with local loss and gain terms, $\gamma_{j,l}= \gamma \delta_{j,l}$, and $P_{j,l}=P\delta_{j,l}$, respectively.
    \item A more complex case, such as the Hatano-Nelson dissipative chain~\cite{Hatano1996,Porras2019,Wanjura2020,Ramos2021}, which requires complex nearest neighbor hopping: $t_{i,j}=\epsilon \delta_{j,l} + t_c (\delta_{i,j+1}e^{i\phi}+\delta_{i,j-1}e^{-i\phi})$, as well as local and non-local decay/pump terms:  $\gamma_{i,j}=\gamma\delta_{ij}+\gamma_{\mathrm{nn}}(\delta_{i,j-1}+\delta_{i,j+1})$ and $P_{i,j}=P\delta_{ij}+P_{\mathrm{nn}}(\delta_{i,j-1}+\delta_{i,j+1})$.
\end{itemize}

In what follows, we show how to use decimation~\cite{Nelson1975,Odashima2016} to obtain the lattice Green's functions, $G_{j,l}(\omega)$, in both models. This technique has been successfully applied in the Hermitian scenario~\cite{Ashraff1988,Chakrabarti1989,Southern1983,Liu1986,Southern1985,Ruocco2017,Pai2019}, and mainly to determine the surface Green's function. However, as will show now, since these results are based on a self-similarity transformation, they can be directly extrapolated to the dissipative scenario and can be used to characterize arbitrary Green's functions.

\subsection{Real space decimation: Method 1~\label{subsec:realspacedecimation}}

In this section, we will see how to determine the self-similarity transformation of the equations of motion in a finite system by projecting Eq.~\eqref{eq:GF-EOM1} to real space. 
Doing that, one can write the following  set of coupled equations for the Green's functions in the steady state
%
\begin{align}
\left(\omega-\tilde{\epsilon}_{0}\right)G_{0,j}\left(\omega\right)=&\delta_{0,j}+\tilde{t}_{+}G_{1,j}\left(\omega\right)\,,\label{eq:EOM1}\\
\left(\omega-\tilde{\epsilon}_{1}\right)G_{1,j}\left(\omega\right)=&\delta_{1,j}+\tilde{t}_{-}G_{0,j}\left(\omega\right)+\tilde{t}_{+}G_{2,j}\left(\omega\right)\,,\label{eq:EOM2}\\
\left(\omega-\tilde{\epsilon}_{2}\right)G_{2,j}\left(\omega\right)=&\delta_{2,j}+\tilde{t}_{-}G_{1,j}\left(\omega\right)+\tilde{t}_{+}G_{3,j}\left(\omega\right)\,,\label{eq:EOM3}\\
&\vdots \nonumber
\end{align}
where $\tilde{\epsilon}_j=t_{j,j}-i(\gamma_{j,j}-P_{j,j})/2$, and $\tilde{t}_{\pm}=t_{j,j\pm 1}-i(\gamma_{j,j\pm 1}-P_{j,j\pm 1})/2$. Note that here we have implicitly assumed that the hopping, decay, and pump terms beyond nearest neighbors vanish, like in the two models of interest for this manuscript. However, if this is not the case, one just needs to consider enlarged unit cells.

To proceed with decimation, we choose to start from the site next to the boundary, i.e., site $1$ (see Fig.~\ref{fig:Schematic}, method 1). This means that we formally solve $G_{1,j}(\omega)$ in Eq.~\eqref{eq:EOM2}, and insert its value in the remaining equations. Notice that different decimation strategies might be more favorable, depending on the Green's function of interest and/or the geometry of the system~\cite{Ashraff1988,Chakrabarti1989,Southern1983,Liu1986,Southern1985,Ruocco2017,Pai2019}. 
After decimating site $1$, we find that the equations of motion can be written as:
\begin{align}
    \left(\omega-\tilde{\epsilon}_{0}^{\prime}\right)G_{0,j}\left(\omega\right)=&\delta_{0,j}^{\prime}+\tilde{t}_{+}^{\prime}G_{2,j}\left(\omega\right)\,,\\\left(\omega-\tilde{\epsilon}_{2}^{\prime}\right)G_{2,j}\left(\omega\right)=&\delta_{2,j}^{\prime}+\tilde{t}_{-}^{\prime}G_{0,j}\left(\omega\right)+\tilde{t}_{+}G_{3,j}\left(\omega\right)\,,\\\left(\omega-\tilde{\epsilon}_{3}\right)G_{3,j}\left(\omega\right)=&\delta_{3,j}+\tilde{t}_{-}G_{2,j}\left(\omega\right)+\tilde{t}_{+}G_{4,j}\left(\omega\right)\,,\\&\vdots \nonumber
\end{align}
where $\tilde{\epsilon}_x^{\prime}=\tilde{t}_{+}\tilde{t}_{-}/(\omega-\tilde{\epsilon}_1)$, $\tilde{t}_{\pm}^{\prime}=\tilde{t}_{\pm}^{2}/(\omega-\tilde{\epsilon}_1)$, $\delta_{0,j}^\prime=\delta_{0,j}+\delta_{1,j}\tilde{t}_{+}/(\omega-\tilde{\epsilon}_1)$ and $\delta_{2,j}^\prime=\delta_{2,j}+\delta_{1,j}\tilde{t}_{-}/(\omega-\tilde{\epsilon}_1)$ are the renormalized parameters after decimating site $1$. Crucially, the renormalized equations are identical to the initial ones, if we re-label sites accordingly. 
This allows us to focus on the transformation rules of the parameters, rather than on the Green's functions themselves.  
After decimating a few more sites, one finds the general transformation rules:
\begin{align}
    \tilde{\epsilon}_{1}^{\left(n+1\right)}&=\tilde{\epsilon}_{n+2}+\frac{\tilde{t}_{-}\tilde{t}_{+}}{\omega-\tilde{\epsilon}_{1}^{\left(n\right)}},\quad
    \tilde{t}_{\pm}^{\left(n+1\right)}=\frac{\tilde{t}_{\pm}^{\left(n\right)}\tilde{t}_{\pm}}{\omega-\tilde{\epsilon}_{1}^{\left(n\right)}}\,,\label{eq:Recurrence}\\
    \tilde{\epsilon}_{0}^{\left(n+1\right)}&=\tilde{\epsilon}_{0}^{\left(n\right)}+\frac{\tilde{t}_{+}^{\left(n\right)}\tilde{t}_{-}^{\left(n\right)}}{\omega-\tilde{\epsilon}_{1}^{\left(n\right)}},\:
    \delta_{0,j}^{\left(n+1\right)}=\delta_{0,j}^{\left(n\right)}+\frac{\tilde{t}_{+}^{\left(n\right)}}{\omega-\tilde{\epsilon}_{1}^{\left(n\right)}}\delta_{1,j}^{\left(n\right)}\,,\\
    \delta_{1,j}^{\left(n+1\right)}&=\delta_{n+2,j}+\frac{\tilde{t}_{-}}{\omega-\tilde{\epsilon}_{1}^{\left(n\right)}}\delta_{1,j}^{\left(n\right)}\,,
\end{align}
where the superscript $(n)$ indicates the number of times that decimation has been applied. These recurrence equations allow us to determine the exact surface Green's function for a system with $N$ sites:
\begin{equation}
    G_{0,j}\left(\omega\right)=\frac{\delta_{0,j}^{\left(N\right)}}{\omega-\tilde{\epsilon}_{0}^{\left(N\right)}}\,,
\end{equation}
Furthermore, the solution can be expressed analytically by noticing that the recurrence equation for $\tilde{\epsilon}_1^{(n+1)}$ can be exactly solved. For example, in the case of homogeneous on-site energies and dissipation (i.e., for $t_{j,j}=\epsilon \to \tilde{\epsilon}_j=\epsilon-i(\gamma-P)/2$):
\begin{equation}
    \tilde{\epsilon}_{1}^{\left(n\right)}=\frac{\left(\tilde{\epsilon}\lambda_{+}+2\right)\lambda_{+}^{n}-\left(\tilde{\epsilon}\lambda_{-}+2\right)\lambda_{-}^{n}}{\lambda_{+}^{n+1}-\lambda_{-}^{n+1}}\,,\label{eq:Solution-Decimation1}
\end{equation}
with
\begin{equation}
    \lambda_{\pm}=\frac{\omega-\tilde{\epsilon}\pm\sqrt{\left(\omega-\tilde{\epsilon}\right)^{2}-4\tilde{t}_{+}\tilde{t}_{-}}}{\tilde{t}_{-}\tilde{t}_{+}}\,.
\end{equation}
From Eq.~\eqref{eq:Solution-Decimation1} it is straightforward to write the solution to all the other recurrence equations, and then, for the surface Green's function $G_{0,j}(\omega)$. 
Importantly, all the other Green's functions can be determined recurrently from $G_{0,j}\left(\omega\right)$ and the original equations of motion [cf.~Eqs.~\eqref{eq:EOM1} to~\eqref{eq:EOM3}]. For example, from Eq.~\eqref{eq:EOM1} we have:
\begin{equation}
    G_{1,j}\left(\omega\right)=\frac{\omega-\tilde{\epsilon}_{0}}{\tilde{t}_{+}}G_{0,j}\left(\omega\right)-\tilde{t}_{+}^{-1}\delta_{0,j}\,.
\end{equation}

\begin{figure}[tb]
    \centering
    \includegraphics[width=1\columnwidth]{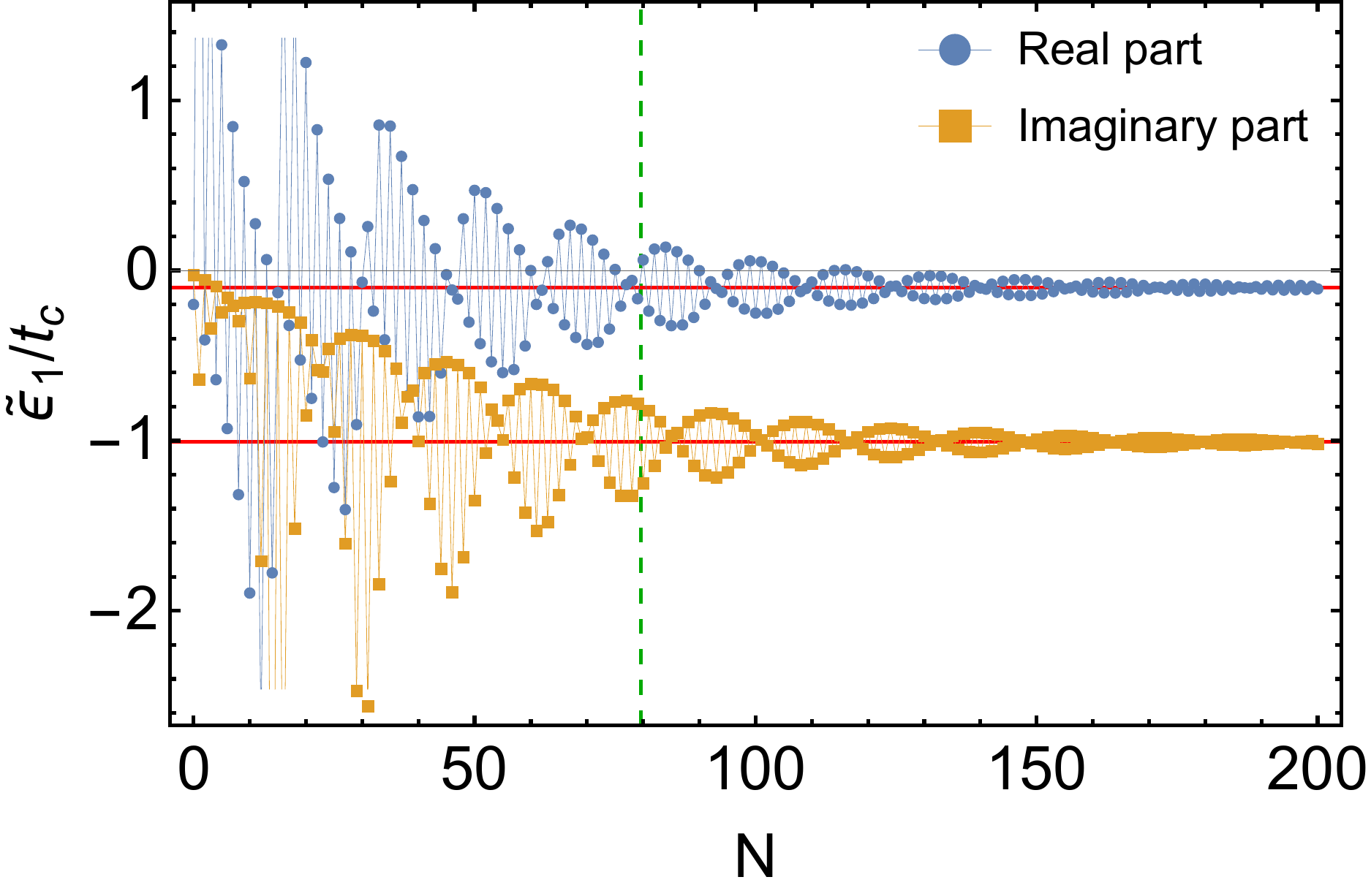}
    \caption{Comparison between the exact value of $\tilde{\epsilon}_1^{N}$ for a finite chain with $N$ sites and the one obtained in the semi-infinite limit (horizontal red lines) for $\gamma/t_c=0.1$, $P/t_c=0.05$, $\epsilon/t_c=-0.2$ and $\omega/t_c=0$. The tendency towards the semi-infinite limit is not monotonous, but is correct for large systems. The vertical dashed line indicates the corresponding correlation length $\Re[\xi(\omega)]^{-1}$ (see Eq.~\eqref{eq:sol-G-simplified} below for details) for these parameters. This length is in one-to-one correspondence with the size required to reach convergence to the semi-infinite limit.}
    \label{fig:Decimation-Check1}
\end{figure}

Although these solutions are analytical for arbitrary $N$, for large systems they are polynomials with large powers and their analysis in simple terms is difficult (although this form is still efficient for numerical calculations).
For this reason we now focus on the semi-infinite limit, $N\to\infty$, where it is possible to find simple analytical expressions which are asymptotically exact. 
This is because in that limit, Eq.~\eqref{eq:Recurrence} becomes an infinite continued fraction:
\begin{equation}
    \tilde{\epsilon}_{1}^{\left(N\right)}\underset{N\to\infty}{\longrightarrow}\tilde{\epsilon}+\frac{\tilde{t}_{-}\tilde{t}_{+}}{\omega-\tilde{\epsilon}-\frac{\tilde{t}_{-}\tilde{t}_{+}}{\omega-\tilde{\epsilon}-\frac{\tilde{t}_{-}\tilde{t}_{+}}{\omega-\tilde{\epsilon}-\ldots}}}\,,
\end{equation}
which can be calculated exactly by re-writing it as a quadratic equation $(\tilde{\epsilon}_{1}^{\left(N\right)}-\tilde{\epsilon})(\omega-\tilde{\epsilon}_{1}^{\left(N\right)})=\tilde{t}_{+}\tilde{t}_{-}$, whose solution is:
\begin{equation}
    \tilde{\epsilon}_{1}^{\left(N\right)}\underset{N\to\infty}{\longrightarrow}\frac{\tilde{\epsilon}+\omega\pm\sqrt{\left(\tilde{\epsilon}-\omega\right)^{2}-4\tilde{t}_{-}\tilde{t}_{+}}}{2}\label{eq:Semi-infinite}\,.
\end{equation}
Here, the sign to get the correct physical solution is fixed by imposing the decay $\tilde{\epsilon}_{1}^{(N)}\to \tilde{\epsilon}$ for $\omega\to\pm\infty$. As a numerical check, in Fig.~\ref{fig:Decimation-Check1} we show a comparison between the exact value of $\tilde{\epsilon}^{(N)}_1$ for a system with $N$ sites and its value in the semi-infinite limit from Eq.~\eqref{eq:Semi-infinite}, showing indeed an excellent agreement. 

In the next subsection, we will show how to arrive to similar expressions for the semi-infinite limit using a different decimation method based on the Dyson's equation.

\subsection{Dyson's equation decimation: Method 2~\label{subsec:dyson}}

Here, we use a slightly different approach consisting in separating $\mathcal{D}$ in two parts: one will be the ``unperturbed" contribution, $\mathcal{D}_0$, which corresponds to $\mathcal{D}$ excluding the hopping to the first site $j=0$. The other part will the ``perturbation", $\mathcal{V}$, introduced by the hopping of the first site to the rest of the chain. 
The advantage of this separation is that, in the semi-infinite limit, the surface Green's function before and after adding the perturbation, coincide.
To see this, notice that after the separation $\mathcal{D}=\mathcal{D}_0+\mathcal{V}$, Eq.~\eqref{eq:GF-EOM1} in the steady state can be expressed as a Dyson's equation:
\begin{equation}
    G(\omega)=g(\omega)+g(\omega)\mathcal{V}G(\omega)\,,\label{eq:Dyson}
\end{equation}
where we have defined the unperturbed dissipative Green's function $g(\omega)=(\omega-\mathcal{D}_0)^{-1}$. For now, the matrix elements of $G(\omega)$ are unknown, and from $g(\omega)$ we just know that the first diagonal element corresponds to the Green's function of the isolated site. In addition, we know that for $g(\omega)$, all the other elements along the first column and first row vanish (because the last site is decoupled). 
Finally, we also know that in the semi-infinite limit $G_{j,l}(\omega)=g_{j+1,l+1}(\omega)$, indicating that the system is large enough so as to remain unchanged with the addition of identical extra sites. Inserting all this knowledge into Eq.~\eqref{eq:Dyson} leads to the fundamental equation for the surface Green's function,  $G_{0,0}\left(\omega\right)$, which reads (see Appendix~\ref{ap:A} for details):
\begin{equation}
    G_{0,0}\left(\omega\right)=g_{0,0}\left(\omega\right)+g_{0,0}\left(\omega\right)\mathcal{V}_{+}G_{0,0}\left(\omega\right)\mathcal{V}_{-}G_{0,0}\left(\omega\right)\label{eq:Eq-G_00}\,,
\end{equation}
where $\mathcal{V}_{\pm}$ represents the complex hopping from the first to the second site, and from the second to the first site, respectively. 
This is one of the fundamental equations for decimation, which shows how to reduce the calculation of the full Green's function to a non-linear equation for the surface Green's function only.

The advantage of this decimation scheme is that Eq.~\eqref{eq:Eq-G_00} is, in general, a non-linear matrix equation, where $g_{0,0}(\omega)$ can describe clusters of sites or even more complex situations such as Bogoliouv-de-Gennes~\cite{Flynn2020,Flynn2021} systems. Interestingly, if the cluster contains a single site, $\mathcal{V}_{\pm}$ coincides with $\tilde{t}_{\pm}$ in Eq.~\eqref{eq:EOM3}. 
Nevertheless, it is not difficult to prove that once the matrix $G_{0,0}(\omega)$ is obtained (either analytically or numerically), the general solution for an arbitrary Green's function matrix is (see Appendix~\ref{ap:A} for details):
\begin{align}
    G_{j,l}=&\left(G_{0,0}\mathcal{V}_{\text{sgn}\left(l-j\right)}\right)^{\left|j-l\right|}G_{0,0}\label{eq:Sol-G}\\
    &+\sum_{a=0}^{\min\left\{j,l\right\}-1}\left(G_{0,0}\mathcal{V}_{-}\right)^{j-a}\left(G_{0,0}\mathcal{V}_{+}\right)^{l-a}G_{0,0}\,,\nonumber
\end{align}
where we have suppressed the $\omega$-dependence for compactness. 
This is an important result, which relates the solution for the surface Green's function in Eq.~\eqref{eq:Eq-G_00}, with an arbitrary Green's function of the system. As a check, from Eq.~\eqref{eq:Sol-G} we have the simple result:
\begin{equation}
    G_{j,0}\left(\omega\right)=\left[G_{0,0}\left(\omega\right)\mathcal{V}_{-}\right]^{j}G_{0,0}\left(\omega\right)\,,\label{eq:simpleG}
\end{equation}
which characterizes the propagation from the edge to site $j$ in terms of the surface Green's function and the complex hopping matrix.
Furthermore, it is useful to separate in Eq.~\eqref{eq:Sol-G} the contribution that depends on the relative distance $d=|j-l|$ and the one that depends on the distance to the surface.
If we assume that $l\geq j$, we can write (the case $l\leq j$ can be analogously obtained) :
\begin{equation}
    G_{j,l}(\omega)=\left[1+\Xi_{j}(\omega)\right]e^{d\xi(\omega)}G_{0,0}\left(\omega\right)\,,\label{eq:sol-G-simplified}
\end{equation}
where the inverse correlation length $\xi(\omega)$ and the matrix $\Xi_j(\omega)$ are given by:
\begin{align}
    \xi(\omega)&=\log\left[ G_{0,0}(\omega)\mathcal{V}_{+}\right],\ \label{eq:Correlation-length}\\
    \Xi_{j}(\omega)&= \sum_{a=0}^{j-1}\left[G_{0,0}\left(\omega\right)\mathcal{V}_{-}\right]^{j-a}\left[G_{0,0}\left(\omega\right)\mathcal{V}_{+}\right]^{j-a}\,.\label{eq:Correlation-length2}
\end{align}
Notice that both $\xi(\omega)$ and $\Xi_j(\omega)$ are matrices of dimension equal to the number of sites in the cluster defining each unit cell. Note also that $|\xi^{-1}(\omega)|$ has units of length, such that when its real part is negative (positive), it defines the exponential decay (amplification) length of the two-point Green's function of the system. This is why along this manuscript we will refer to $\Re[\xi(\omega)]^{-1}$ as the coherence or correlation length of the system.

\subsection{Transient dynamics~\label{subsec:dynamics}}

After having reviewed two methods to obtain the steady-state properties of the system, let us now discuss another application of decimation to quantum dissipative systems which has been largely overlooked: the possibility to efficiently study the transient dynamics and extract analytical expressions for the characteristic times. This is important because certain dissipative systems might have divergent steady state solutions, i.e., those which are dynamically unstable, but still be interesting during their transient dynamics.

To extract the dynamics it is useful to work with the Laplace transform, rather than the Fourier transform that we used to characterize steady state properties. The main difference with the Fourier transform is that the boundary condition for the Laplace transform requires fixing an initial time.  Importantly, the Laplace transform of the equations still allows to apply our decimation scheme, and thus, by means of the inverse Laplace transform, study the transient dynamics and extract certain characteristic times.

To see the role of the Laplace transform of the Green's function, $G(s)=(i s-\mathcal{D})^{-1}$, notice that it solves the equation of motion for the field operator:
\begin{equation}
   \langle a_j(s)\rangle=i \sum_{l}G_{j,l}(s)\langle a_l(0)\rangle\,, \label{eq:Laplace-Transform}
\end{equation}
with $\langle a_j(0) \rangle$ being the field operator at some initial time $t=0$, and $\langle a_i(s) \rangle$ the Laplace transform of the bosonic operator. Notice that since we are not considering any input field/coherent drive, a non-trivial solution of Eq.~\eqref{eq:Laplace-Transform} can only be obtained if we start with an initial seed $\langle a_j (0)\rangle \neq0$.

To find $G(s)$ we just need to realize that the Laplace transform of Eq.~\eqref{eq:GF-EOM1} gives a similar structure as the Fourier transform.
Therefore, the decimation process can be carried out analogously, ending up with the same fundamental equation to determine the surface Green's function [cf Eq.~\eqref{eq:Eq-G_00}]. For all practical purposes, it means that we can just make the substitution $G_{0,0}(\omega\to is)$ to determine the Laplace transform.
However, the main difficulty to study the transient dynamics relies in calculating the inverse Laplace transform. Although there are several numerical methods to deal with this problem~\cite{Kuhlman2012}, here, we present a different approach that allows us to find an exact analytical expression for the Green's function, which is valid for arbitrary 1D dissipative quantum systems with a single site per unit cell.

We start by considering the equation for the surface Green's function in the semi-infinite limit of a dissipative quantum system:
\begin{equation}
    G_{0,0}\left(s\right)=g_{0,0}\left(s\right)+g_{0,0}\left(s\right)\mathcal{V}_{+}G_{0,0}\left(s\right)\mathcal{V}_{-}G_{0,0}\left(s\right)\,,
\end{equation}
As aforementioned, the only difference with Eq.~\eqref{eq:Eq-G_00} is in the definition of the unperturbed one: $ g_{0,0}(s)=\left(is-\mathcal{D}_0\right)^{-1}$. Therefore, the Laplace transform of Green's function terms $G_{i,j}(s)$ fulfils an identical solution as in Eq.~\eqref{eq:Sol-G}.

Now let us focus on the Green's function $G_{j,0}(s)=\left[  G_{0,0}(s) \mathcal{V}_{-} \right]^{j}G_{0,0}(s)$, which controls the propagation from the edge, although the general case $G_{j,l}(s)$ can be analogously treated because its $s$-dependence is analogous. If we focus on the case of one site per unit cell, we conclude that we need the inverse Laplace transform of $G_{0,0}(s)^{j+1}$ to fully determine the time evolution. Typically, the inverse Laplace transform of non-linear functions is hard to find, but in this case, the use of the binomial series allows us to reduce it to the calculation of the inverse Laplace transform of the unperturbed Green's function $g_{0,0}(s)$, which can be done analytically. 

The solution for the surface Green's function for a bosonic chain described by Eqs.~\eqref{eq:H}-\eqref{eq:pump} is:
\begin{equation}
G_{0,0}\left(s\right)=\frac{1-\sqrt{1-4g_{0,0}\left(s\right)^{2}\tilde{t}_{+}\tilde{t}_{-}}}{2g_{0,0}\left(s\right)\tilde{t}_{+}\tilde{t}_{-}}\,.
\end{equation}
We define $2g_{0,0}\left(s\right)=y\left(s\right)$ and $z\left(y\right)=\sqrt{1-y^{2}\alpha}$,
being $\alpha=\tilde{t}_{+}\tilde{t}_{-}.$ This allows us to write
the required expression as:
\begin{align}
G_{0,0}\left(s\right)^{j+1} = & \sum_{n,p=0}^{\infty}\left(\begin{array}{c}
j+1\\
n
\end{array}\right)\left(\begin{array}{c}
\frac{n}{2}\\
p
\end{array}\right)\left(-1\right)^{n+p}\nonumber\\
&\times\alpha^{p-j-1}y^{2p-j-1}\,,
\end{align}
where $\left(\begin{array}{c}
a\\
b
\end{array}\right)$ are the generalized binomial coefficients, and we have applied a binomial series expansion to $\left(1-z\right)^{j+1}$ and to $\sqrt{1-y^{2}\alpha}$.
Now it is possible to perform the inverse Laplace transform of each term in the sum:
\begin{equation}
\mathcal{L}^{-1}\left\{ y\left(s\right)^{2p-j-1}\right\} \left(t\right)=-i2^{2p-j-1}\frac{e^{-it\tilde{\epsilon}}\left(-it\right)^{2p-j-2}}{\Gamma\left(2p-j-1\right)}\,,
\end{equation}
being $\Gamma(x)$ the Gamma function. Now, summing over $p$ and $n$ we can find the analytical expression for $G_{0,0}\left(t\right)^{j+1}$ (see Appendix~\ref{ap:C} for details)
\begin{align}
G_{0,0}\left(t\right)^{j+1} & =e^{-it\tilde{\epsilon}}\left(-i\right)^{j+1}\left(j+1\right)\frac{\mathcal{J}_{j+1}\left(2t\sqrt{\alpha}\right)}{t\alpha^{\frac{j+1}{2}}}\,.\label{eq:Exact-Transient}
\end{align}
Here, $\mathcal{J}_{r}\left(x\right)$ is the $r$-th Bessel function of the first kind, and $\tilde{\epsilon}$ contains the on-site energy and the local dissipation rates, as defined below Eq.~\eqref{eq:EOM3}.

The exact expression in Eq.~\eqref{eq:Exact-Transient} allows us to express, for example, the exact time-evolution $\langle a_j(t) \rangle$ for a coherent state initially prepared at the edge of the chain with amplitude $\langle a_{j}(0) \rangle = \alpha_0 \delta_{j,0}$, which reads:
\begin{align}
\langle a_{j}\left(t\right)\rangle=&i \alpha_0 e^{-it\tilde{\epsilon}}\left(-i\right)^{j+1}\left(j+1\right)\left(\frac{\tilde{t}_{-}}{\tilde{t}_{+}}\right)^{\frac{j}{2}}\nonumber\\
&\times \frac{\mathcal{J}_{j+1}\left(2t\sqrt{\tilde{t}_{+}\tilde{t}_{-}}\right)}{t\sqrt{\tilde{t}_{+}\tilde{t}_{-}}}\,.\label{eq:CoherentStateEvol}
\end{align}

The advantage of this analytical expression is that it allows to capture some general features of the transient dynamics of one-dimensional systems such as the long time behavior or the characteristic velocity of a wave packet.
Also, the steady state solution can simply be obtained from: $\lim_{t\to\infty}f\left(t\right)=\lim_{s\to0}s\mathcal{L}\left\{ f\left(t\right)\right\} \left(s\right)$. This provides a direct way to check the stability of the system.

\section{Examples: driven-dissipative bosonic chains~\label{sec:examples}}

Now, we will apply the tools developed in the previous section to the two particular examples described below Eqs.~\eqref{eq:H}-\eqref{eq:pump}, that are, a coupled-cavity array with individual decay and pump terms (in~\ref{subsec:simple1D}), and the Hatano-Nelson model (in~\ref{subsec:hatano}).

\subsection{Dissipative coupled-cavity arrays~\label{subsec:simple1D}}

We start by analyzing the simplest instance of a driven-dissipative bosonic chain: a coupled-cavity array with nearest neighbor hopping, $t_c$, and individual decay rate, $\gamma$, and pumping, $P$. This is useful, because it will allow us to connect with physics of one-dimensional bosonic chains without dissipation, which has been thoroughly studied in the literature~\cite{lekien05a,longui06a,Longhi2006,garmon13a,calajo16a,sanchezburillo17a,Gonzalez-Tudela2017a}. Note, that the dissipative case without gain have also been considered in other works, e.g., Ref.~\cite{calajo16a}.

\begin{figure}
    \centering
    \includegraphics[width=1\columnwidth]{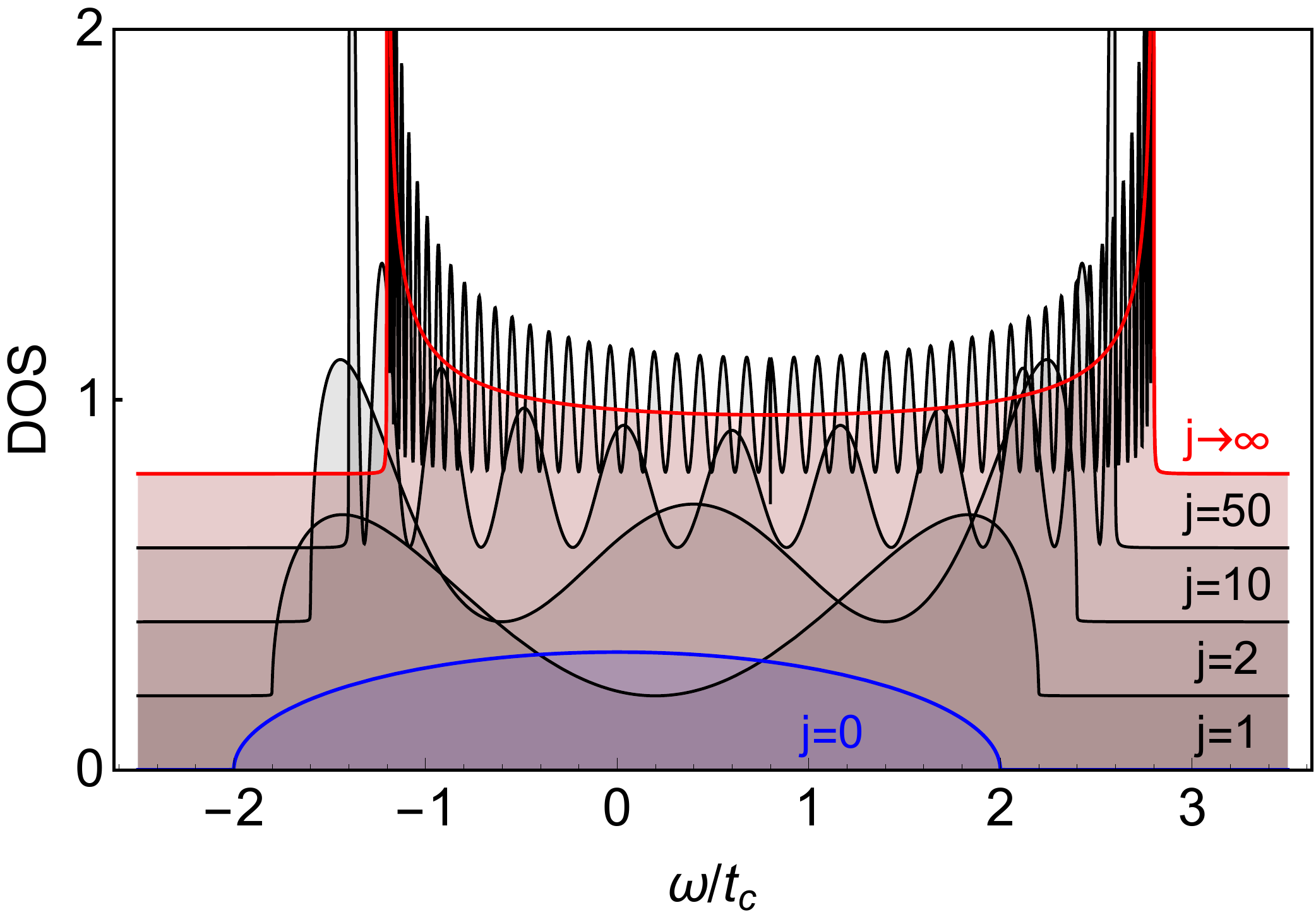}
    \caption{Change in the local density of states (DOS), $D_j(\omega)$, obtained from the exact Green's function from decimation, Eq.~\eqref{eq:sol-G-simplified}, as one moves away from the edge (blue) towards the bulk (red) (intermediate sites: $1$,$2$,$10$ and $50$). Note, the renormalization of the band-edge Van-Hove singularities appearing in the bulk when the position becomes closer to the edge.}
    \label{fig:EvolutionDos}
\end{figure}

Let us first describe a standard result for periodic boundary conditions (PBC) in the infinite limit.
There, the Hamiltonian of this model can be immediately diagonalized:
\begin{align}
H_{\mathrm{1d}}=\sum_k \omega_k a_k^\dagger a_k\,,
\end{align}
with $\omega_k=\omega_a-2 t_c\cos(k)$, being $\omega_a$ the on-site energy of the cavities. 
This expression allows one to find a compact analytical expression for the two-point Green's Function in the limit of $N\rightarrow\infty$ and for $\omega-\omega_a\leq 2 t_c$ given by (see Appendix~\ref{ap:B} for details):
\begin{equation}
    G_{j,l}\left(\omega\right)= -i\frac{e^{\left|j-l\right|\log \left(\frac{\omega-\omega_{a}}{2t_c}- i\sqrt{1-\left(\frac{\omega-\omega_{a}}{2t_c}\right)^{2}}\right)}}{\sqrt{\left(\omega-\omega_{a}\right)^{2}-4t_c^{2}}} \,.\label{eq:Gijbulk}
\end{equation}
From this equation, we can immediately see some of the well-known properties for the case of lossless and infinite bosonic chains, that are: i) the divergent behaviour of the local density of states, $D_j(\omega)=-\frac{1}{\pi}\Im G_{j,j}(\omega+i\eta)$~\footnote{In the lossless bosonic chain scenario $\eta$ is a small imaginary number introduced to obtain a smooth representation} around the band-edges, i.e., $|\omega-\omega_a|\sim 2 t_c$; ii) the divergence of the coherence length, $\Re[\xi(\omega)]^{-1}$, defined in Eq.~\eqref{eq:sol-G-simplified}, for energies within the band, which translates into an infinite range two-point Green's function~\cite{lekien05a,longui06a,Longhi2006,garmon13a,calajo16a,sanchezburillo17a,Gonzalez-Tudela2017a}. 
As we show next, these two features change dramatically when both finite size effects and loss/pump terms are included in the model.

In order to illustrate it, we apply the decimation technique described in Section~\ref{sec:decimation} to obtain $G_{j,l}(\omega)$ for a dissipative, semi-infinite chain. Let us first illustrate the renormalization of the local density of states due to surface effects, as shown in Fig.~\ref{fig:EvolutionDos}, where, we plot the change in the local density of states (DOS) as one moves away from the edge (blue) towards the bulk (red). There, we observe how the initial density of states at the edge develops a number of nodes equal to the number of sites from the edge, and slowly forms the Van-Hove singularities typically observed in the bulk by accumulating states at the band edge. Importantly, it is also possible to obtain the analytical expression for the Green's function with PBC [i.e., to recover Eq.~\eqref{eq:Gijbulk}], by just coupling two semi-infinite chains using Dyson's equation.
\begin{figure}
    \centering
    \includegraphics[width=1\columnwidth]{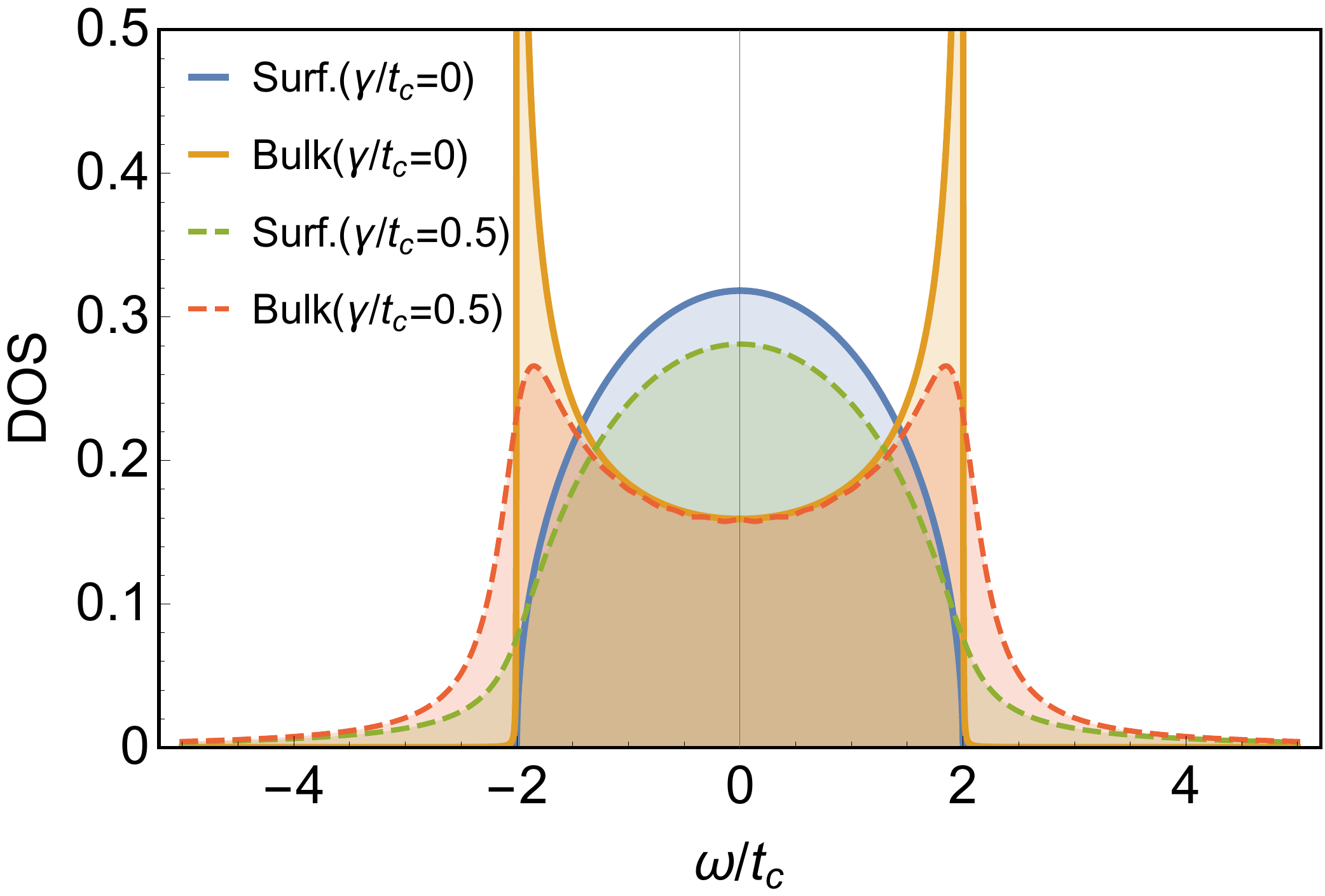}
    \caption{Density of states (DOS), $D_j(\omega)$, as defined in the text, at the boundary (blue and green) and deep in the bulk (orange and red). Solid lines refer to the lossless case, characterized by a sharp disappearance of states beyond $|\omega|\geq 2t_c$ and the Van-Hove singularities for the bulk case. In contrast, the dissipative case softens and renormalizes the singularities, extending the surface/bulk DOS to frequencies beyond $|\omega|=2t_c$.}
    \label{fig:Surface&Bulk-DOS1}
\end{figure}

Let us now focus on the effect of dissipation on such local density of states. This is shown in Fig.~\ref{fig:Surface&Bulk-DOS1}, where we compare the DOS for the surface site and for a site in the bulk with and without dissipation. One can see that the surface DOS reproduces the well known profile of the Hermitian case (solid blue), while the presence of dissipation transfers spectral weight to higher/lower energy states, with $|\omega|/t_c>2$ (green dashed), by softening the band-edges . Analogously, the DOS in the bulk reproduces the well-known Van-Hove singularities at $|\omega|/t_c=2$ (solid yellow) and the presence of dissipation again softens their presence and populates higher/lower energy states. 

\begin{figure}
    \centering
    \includegraphics[width=1\columnwidth]{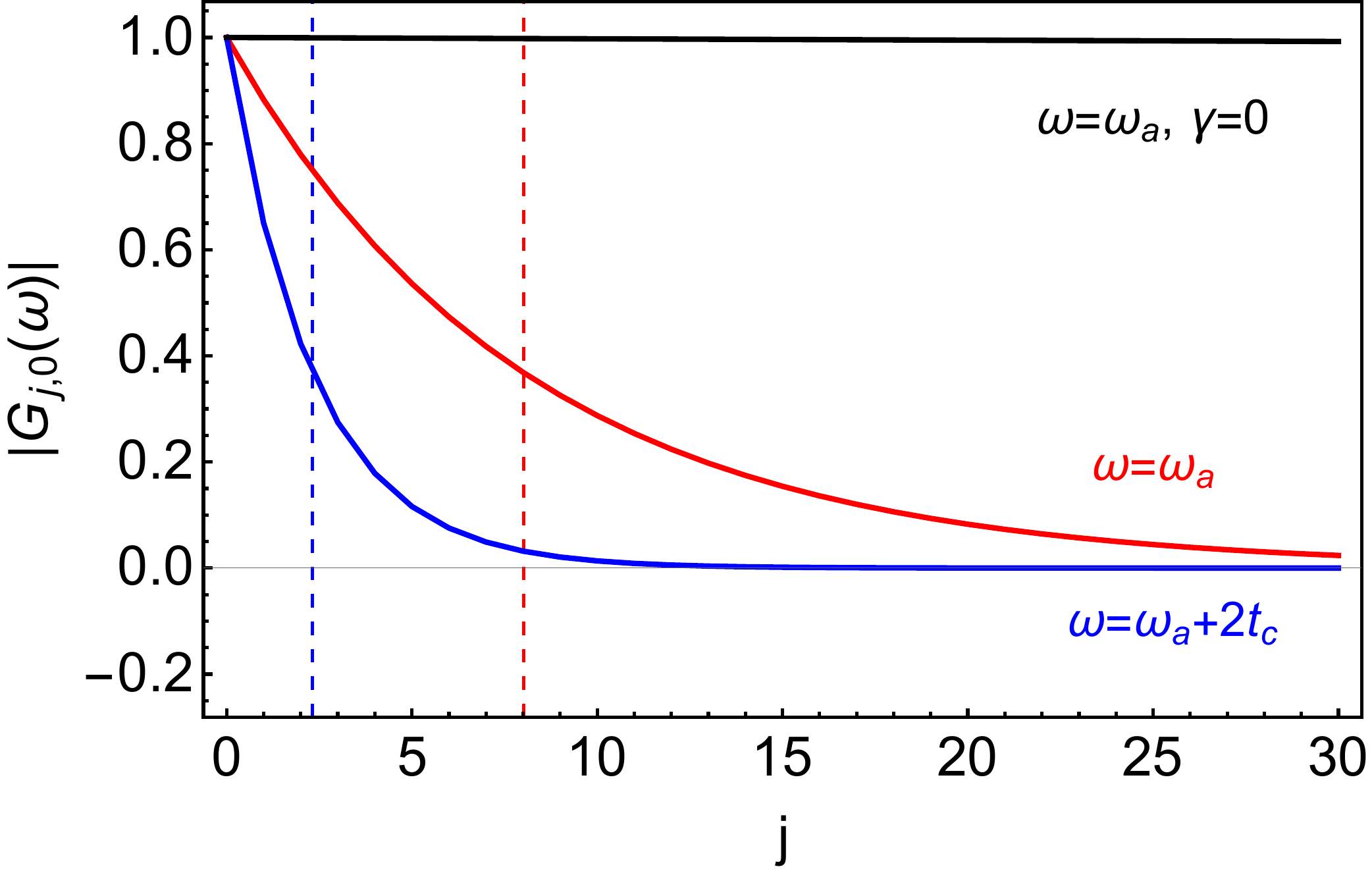}
    \caption{$|G_{j,0}(\omega)|$ normalized by $|G_{0,0}(\omega)|$ for $\omega=\omega_a$(red), $\omega=\omega_a+2t_c$ (blue) and $\omega=\omega_a$ with $\gamma/t_c=0$ (black). In the lossless case, 2-point correlations remain constant between distant sites. In contrast, adding dissipation introduces a decay of correlations, which is accentuated at the band edges. Vertical lines indicate the corresponding coherence length $1/\Re [\xi(\omega)]$ from Eq.~\eqref{eq:sol-G-simplified}.}
    \label{fig:Decay&CoherenceLength}
\end{figure}

Finally, let us illustrate the impact of dissipation on the two-point Green's functions. As aforementioned, in lossless 1D systems correlations between sites do not decay with the distance.  As expected, dissipation introduces a decay, which can be characterized with our analytical expressions. In Fig.~\ref{fig:Decay&CoherenceLength} we plot the absolute value of the 2-point Green's function as a function of the distance, normalized by its value at the surface.
As expected, the lossless case (black) shows constant correlation between sites, independently of the distance, for states within the band (otherwise there is damping). 
When dissipation is added, even states within the band display a decay (red), which is accentuated if one moves away from the band center (blue).

These changes can be systematically studied in terms of the inverse correlation length $\xi(\omega)$, which can be written as:
\begin{equation}
    \xi(\omega)=\frac{1}{2}\log\left(\frac{\tilde{t}_{-}}{\tilde{t}_{+}}\right)+\log\left[\alpha\left(\omega\right)-\alpha\left(\omega\right)\sqrt{1-\alpha\left(\omega\right)^{-2}}\right],\label{eq:xi-decay}
\end{equation}
with $\alpha(\omega)=(\omega-\omega_a+i\Gamma)/2\sqrt{\tilde{t}_{+} \tilde{t}_{-}}$, and $\tilde{t}_{\pm}=t_{j,j\pm 1}-i(\gamma_{j,j\pm 1}-P_{j,j\pm 1})/2$. In addition, $\Gamma=\gamma-P$ denotes the net loss/gain rate. In particular, for a dissipative chain with local gain and loss, the first term in Eq.~\eqref{eq:xi-decay} vanishes because $\tilde{t}_\pm=t_c$~\footnote{Note, this will not be the case in the Hatano-Nelson model below, where the non-reciprocal nature of the model is reflected in that term}. Thus, in this case one just needs to study the second term, which has two distinct regimes: (i) $|\alpha(\omega)|\gg 1$ or highly dissipative and (ii) $|\alpha(\omega)|\ll 1$ or weakly dissipative.
In the weakly dissipative case we can expand to first order in $\Gamma$ to find:
\begin{equation}
    \xi(\omega)\simeq \xi_0(\omega)-\frac{\Gamma}{\sqrt{4t_c^2-(\omega-\omega_a)^2}},
\end{equation}
being $\xi_0(\omega)$ the inverse coherence length in absence of dissipation. This shows that in the presence of weak dissipation, the real part of the correlation length is $\Re[\xi(\omega)]^{-1}\simeq -\Gamma/\sqrt{4t_c^2-(\omega-\omega_a)^2}$, indicating that correlations between sites decay exponentially with the distance as expected: the larger the net loss ($\Gamma$) and/or the smaller the group velocity ($\sqrt{4t_c^2-(\omega-\omega_a)^2}$), the stronger is the exponential damping. Besides, an important observation is that since what enters in the equations is the net loss/gain rate, $\Gamma$, one can compensate the exponential damping of the two-point Green's function by adding gain to the system, until $\Gamma=0$, where the infinite correlation length of the lossless 1D systems is recovered.

Finally, we briefly discuss the asymptotic behavior of the transient dynamics for a reciprocal chain with local dissipation only.
From Eq.~\eqref{eq:CoherentStateEvol} we can see that it particularizes to the following expression:
\begin{equation}
\langle a_{j}\left(t\right)\rangle = i\alpha_0 e^{-i\tilde{\epsilon}t}\left(-i\right)^{j+1}\left(j+1\right)\frac{\mathcal{J}_{j+1}\left(2t_c t\right)}{t_c t}.
\end{equation}
Then, if we are interested in the behaviour at long time, we can use the asymptotic expansion of the Bessel functions $\mathcal{J}_{m+1}\left(2t_c t\right)\sim\left(\pi t_c t \right)^{-1/2}$, to find the well-known $t^{-3/2}$ behavior:
\begin{equation}
\langle a_{j}\left(t\right)\rangle \sim i\alpha_0 e^{-i\tilde{\epsilon}t}\left(-i\right)^{j+1}\frac{j+1}{\sqrt{\pi}\left(t_c t\right)^{3/2}},
\end{equation}
where we have neglected the oscillating part of the expansion, as in experiments one typically is interested in the average value.
This behaviour perfectly captures the dynamics at long time and agrees with previous works where the asymptotic limit has also been obtained~\cite{longui06a,Longhi2006,garmon13a,sanchezburillo17a,Gonzalez-Tudela2017a,Garmon2019Non-MarkovianContinuum}.

\subsection{Hatano-Nelson chain~\label{subsec:hatano}}

Now, let us consider a more complex model, that is the Hatano-Nelson~\cite{Hatano1996}, which is a paradigm of topological quantum amplifiers induced by dissipation~\cite{Porras2019,Wanjura2020,Ramos2021,Gomez-Leon2021}. The minimal instance of this model requires complex nearest-neighbour hoppings, $t_c e^{i\phi}$, local dissipative/pump terms, $\gamma$ and $P$, and non-local nearest-neighbour decay or pumping ($\gamma_{\mathrm{nn}}$ and $P_{\mathrm{nn}}$, respectively).
Here we choose the presence of non-local gain $P_{nn}\neq0$, and hence, we can set $\gamma_{nn}=0$ for simplicity.
Furthermore, in the standard form of this model, local and non-local gain are not independent, as they are related by $P=2P_{\mathrm{nn}}$~\cite{Gomez-Leon2021,Ramos2021}. 
As shown in several works~\cite{Porras2019,Wanjura2020,Gomez-Leon2021}, this model supports a topological amplifying phase where excitations propagate along one direction only, exponentially increasing their particle number with the distance of propagation.

Applying the results obtained by decimation in Section~\ref{sec:decimation}, we obtain that the surface Green's function for an isolated site is given by:
\begin{equation}
    g_{0,0}(\omega)=\frac{1}{\omega-\epsilon+i\frac{\gamma-P}{2}}.
\end{equation} 
The calculation of the surface Green's function requires to solve Eq.~\eqref{eq:Eq-G_00}, which for the present model is just a second order equation with solution:
\begin{equation}
    G_{0,0}(\omega)=\frac{1-\sqrt{1-4 g_{0,0}(\omega)^2 \tilde{t}_{+}\tilde{t}_{-}}}{2 g_{0,0}(\omega) \tilde{t}_{+}\tilde{t}_{-}},\label{eq:GF-example0}
\end{equation}
with $\tilde{t}_{\pm}=t_c e^{\pm i\phi}+iP/4$. The analytical expression for the surface Green's function allows us to calculate the two relevant quantities that define the Green's function in Eq.~\eqref{eq:sol-G-simplified}:
\begin{align}
    \xi(\omega)=&\log \left[G_{0,0}(\omega)\tilde{t}_{+}\right]\,,\label{eq:Hatano-xi}\\
    \Xi_j(\omega)=&\rho(\omega)\frac{\rho(\omega)^j-1}{\rho(\omega)-1}\,,
\end{align}
that are the inverse correlation length $\xi(\omega)$ and site-dependent amplitude $\Xi_j(\omega)$, respectively. Note, that we have evaluated the sum over $a$ in Eq.~\eqref{eq:Correlation-length2} and that $\rho(\omega)=G_{0,0}\left(\omega\right)^{2}\tilde{t}_{-}\tilde{t}_{+}$. Therefore, we can write the Green's function in the Hatano-Nelson model with the following compact expression:
\begin{equation}
    G_{j,l}\left(\omega\right)=\frac{e^{\left(j+1\right)\log\left[\rho\left(\omega\right)\right]}-1}{\rho\left(\omega\right)-1}e^{\left(l-j\right)\xi\left(\omega\right)}G_{0,0}\left(\omega\right)\label{eq:semi}\,.
\end{equation}
\begin{figure}[tb]
    \centering
    \includegraphics[width=1\columnwidth]{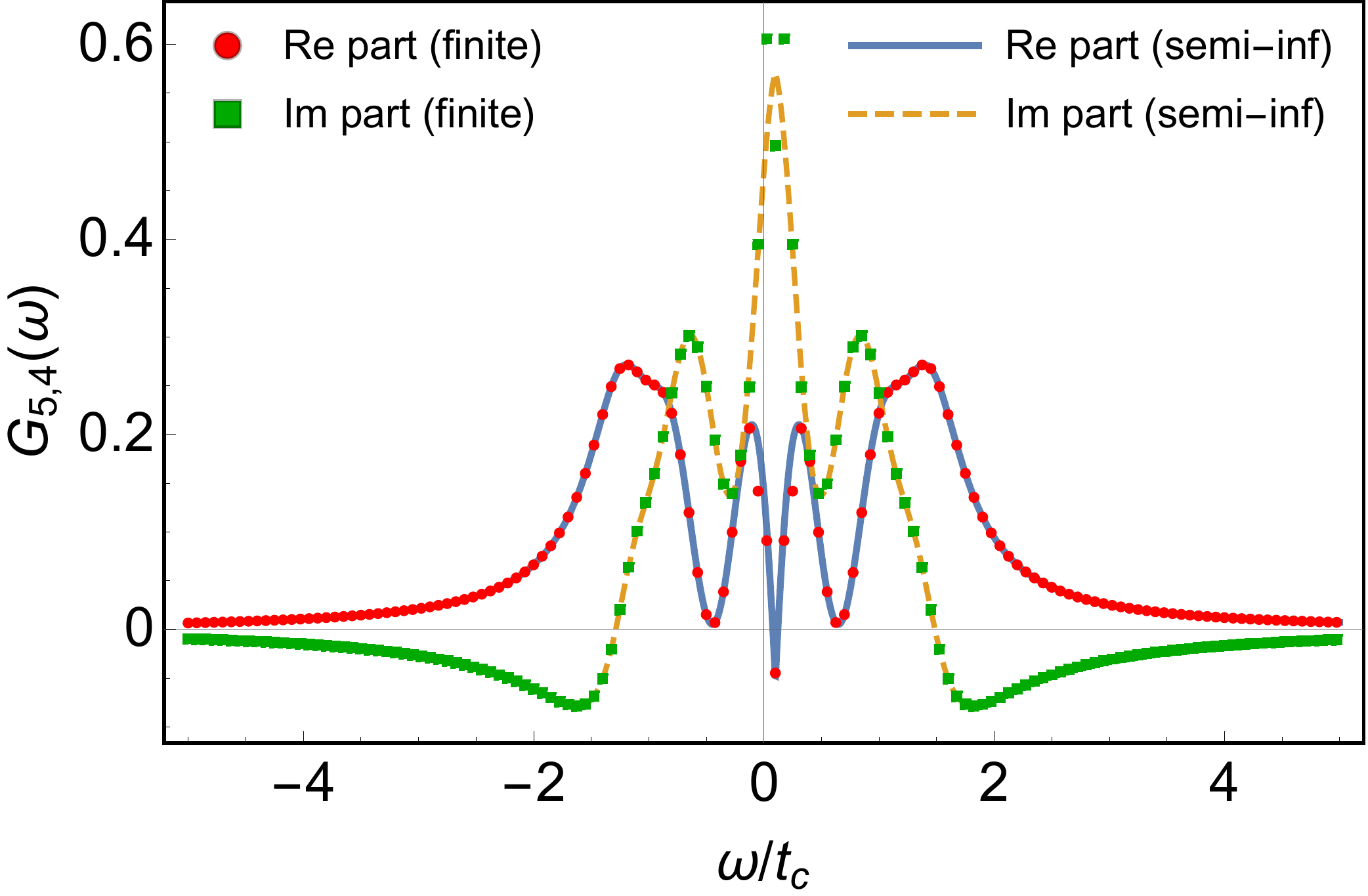}
    \caption{Comparison between $G_{5,4}(\omega)$ from exact diagonalization for a finite array with $N=45$ sites (dot and square markers) and the semi-infinite case (solid and dashed lines), computed using Eq.~\eqref{eq:semi}. Parameters: $\gamma/t_c=3$, $P/t_c=3$, $\epsilon/t_c=0.1$ and $\phi=0.9\pi/2$.}
    \label{fig:G-check-HN}
\end{figure}

\begin{figure}[tb]
    \centering
    \includegraphics[width=1\columnwidth]{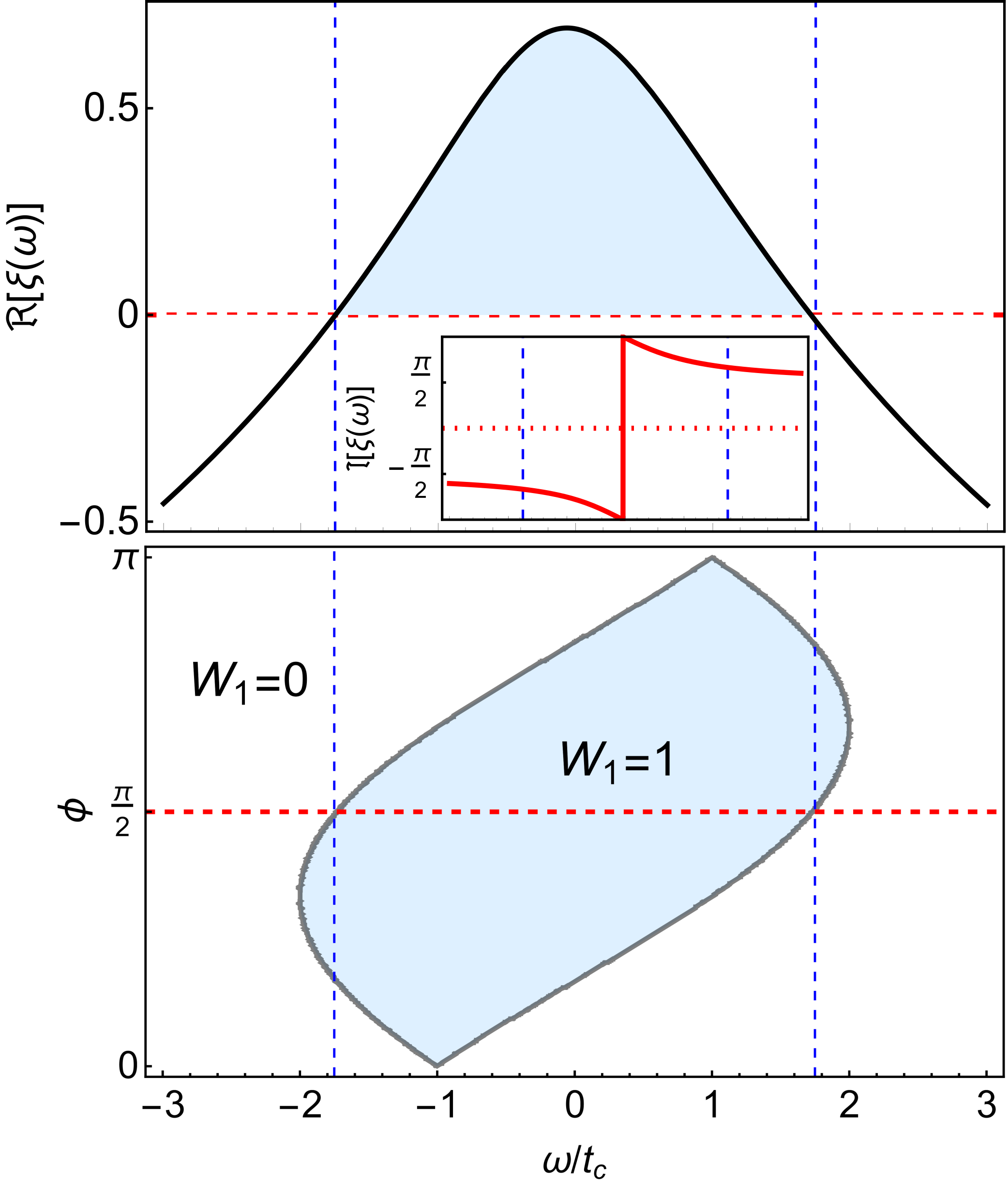}
    \caption{(Top) $\xi(\omega)$ for the topological phase ($\phi=\pi/2$). The blue region represents the region of amplification due to $\Re [\xi(\omega)]>0$. The vertical dashed lines indicate the critical points from the phase diagram. The inset shows the $\Im [\xi(\omega)]$.
    (Bottom) Phase diagram obtained from the surface Green's function. It gives identical results to the calculation of $W_1(\omega)$ for PBC using Eq.~\eqref{eq:Winding-Contour}. Parameters: $\gamma/t_c=2$, $P/t_c=4$ and $\epsilon/t_c=0$.}
    \label{fig:Xi-HN}
\end{figure}

To validate this expression obtained through decimation, in Fig.~\ref{fig:G-check-HN} we show a comparison between the Green's function obtained from exact diagonalization for the particular component $G_{5,4}(\omega)$ (for a finite chain with $N=45$ sites), and the semi-infinite solution obtained in Eq.~\eqref{eq:semi}. There, it can be seen how the agreement between the two is excellent for the choice of parameters depicted in the caption. Although not shown, other parameters show similar agreement, as long as $N$ is large enough.

In contrast with the case of a simple dissipative chain studied in~\ref{subsec:simple1D}, the combination of collective dissipation, non-reciprocity and time-reversal symmetry breaking due to the gauge field produce a topological amplification phase \cite{Porras2019,Wanjura2020,Ramos2021,Gomez-Leon2021}. 
As the amplification process depends on the distance $|x-y|$ traveled by an excitation, it should be encoded in the real part of the inverse correlation length $\xi(\omega)$, which can be easily computed from Eq.~\eqref{eq:Hatano-xi}.  
Fig.~\ref{fig:Xi-HN}~(top) shows its behavior as a function of $\omega$ and the blue area shows the region $\Re[\xi(\omega)]>0$, where amplification happens. In addition, the inset shows the behavior of the imaginary part of $\xi(\omega)$, which is related to the phase acquired by the excitations at a given energy. The vertical~(blue, dashed) gridlines in Fig.~\ref{fig:Xi-HN}~(top) correspond to the phase boundaries obtained when $\Re[\xi(\omega)]$ changes its sign. 
Interestingly, these boundaries coincide with the ones of the topological phase diagram for periodic boundary conditions (Fig.~\ref{fig:Xi-HN}, bottom), which in this case can be calculated as~\cite{Gomez-Leon2021}:
\begin{equation}
    W_{1}\left(\omega\right)=\int_{-\pi}^{\pi}\frac{dk}{2\pi i}\partial_{k}\log\left[\omega-\mathcal{D}\left(k\right)\right],
\end{equation}
with $\mathcal{D}\left(k\right)=\epsilon-i\frac{\gamma}{2}+iP\cos^{2}\left(\frac{k}{2}\right)+2t_{c}\cos\left(k-\phi\right)$. Using $z=e^{ik}$, one can calculate this integral analytically resulting in:
\begin{equation}
W_{1}\left(\omega\right)=1-\theta\left(1-\left|z_{+}\right|\right)-\theta\left(1-\left|z_{-}\right|\right),\label{eq:Winding-Contour}
\end{equation}
with $z_{\pm}=\left(\mu\pm\sqrt{\mu^{2}-4\tilde{t}_{+}\tilde{t}_{-}}\right)/2\tilde{t}_{-}$ being the poles of the integrand and $\mu=\omega-\epsilon+i\left(\gamma-P\right)/2$. 
The exact agreement between $W_1(\omega)$ and the positive regions of the coherence length in Fig.~\ref{fig:Xi-HN} (bottom), allows us to postulate that $\Theta(\Re\left[ \xi(\omega)\right])$ can be used as topological invariant for this model. 
This is important because calculating the inverse correlation length $\xi(\omega)$ only requires to know the surface Green's function $G_{0,0}(\omega)$, and it can be calculated analytically in the semi-infinite system, numerically for a finite system or from a local measurement in an experimental setup, allowing us to link a local observable with the topology of the model.



\subsubsection{Transient dynamics in the Hatano-Nelson chain~\label{subsec:amplification}}

We now go beyond the steady-state properties and we study the dynamics that can be extracted from the expressions obtained in Section~\ref{subsec:dynamics}, starting by an initial state situation, e.g., where one of the edges is populated.

\begin{figure}[tb]
    \centering
    \includegraphics[width=1\columnwidth]{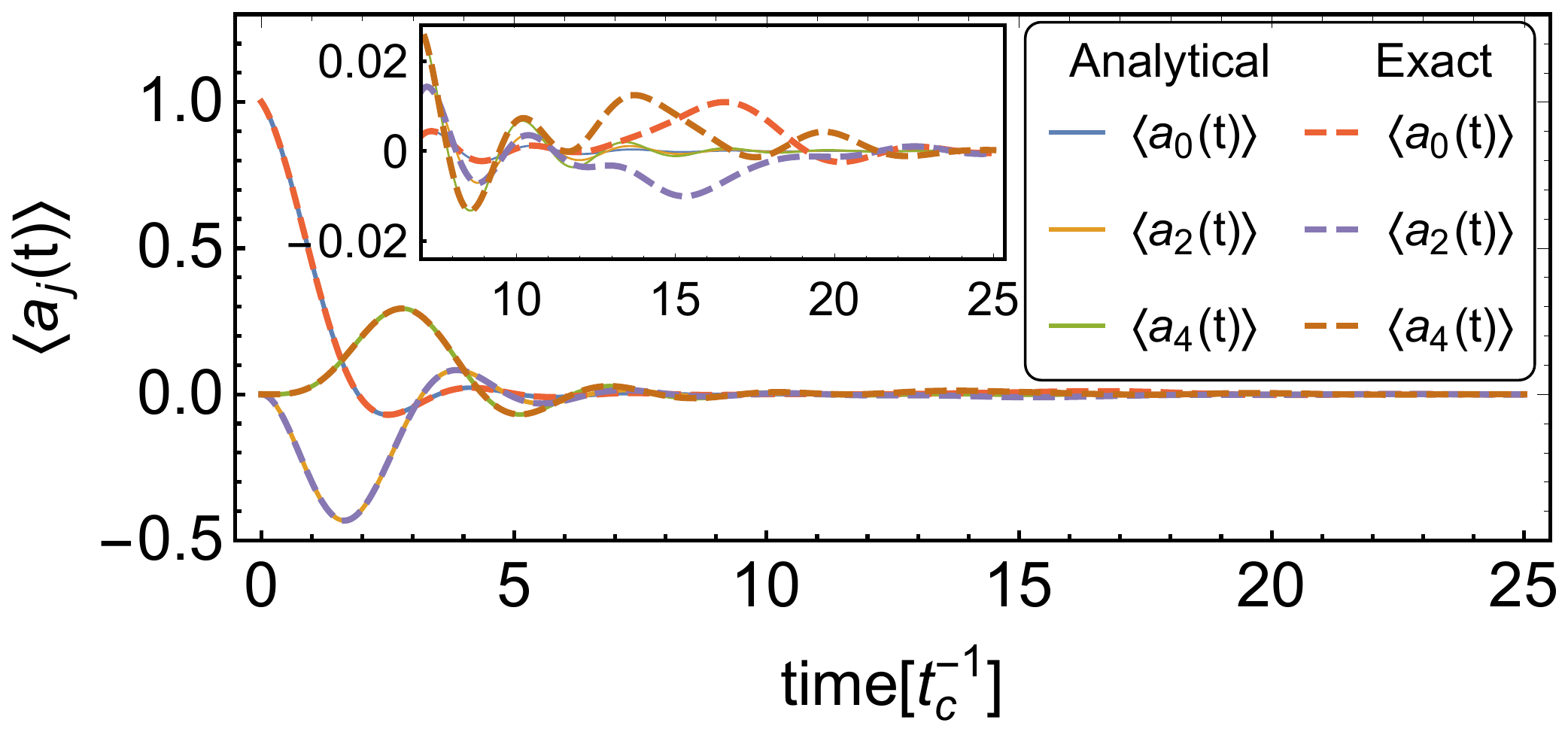}
    \caption{Comparison between the exact and the analytical solution for the real part of $\langle a_{j}(t)\rangle$ for $N=15$, $\phi=0$, $P/t_c=0$, $\gamma/t_c=0.5$ and $\epsilon/t_c=0.1$. The inset shows a zoom at intermediate time, where a difference between the two solutions arises around $t\sim 12 t_c^{-1}$ due to a revival. This is the Poincar\'e recurrence time for the finite system, where the excitation has bounced off the opposite edge. Enlarging the unit cell size delays this effect.}
    \label{fig:Transient1}
\end{figure}

In the Hatano-Nelson model, there are two well-distinguished regimes. One can choose the parameters to be in the topologically trivial phase, dominated by losses. In that case, the initial coherent state propagates and decays over time, as shown in Fig.~\ref{fig:Transient1}, as it would occur in a standard dissipative chain like the one studied in Section~\ref{subsec:simple1D}. There, one can see how the excitation amplitude is damped as it moves through the array. The comparison shows excellent agreement between the exact numerical solution for a finite size system and the semi-infinite analytical expression. Interestingly, the inset shows that finite size effects can emerge for long enough time, due to the Poincar\'e recurrence time. Therefore, our solutions allow us to separate interference due to the opposite boundary from the bulk dynamics.

\begin{figure}[tb]
    \centering
    \includegraphics[width=1\columnwidth]{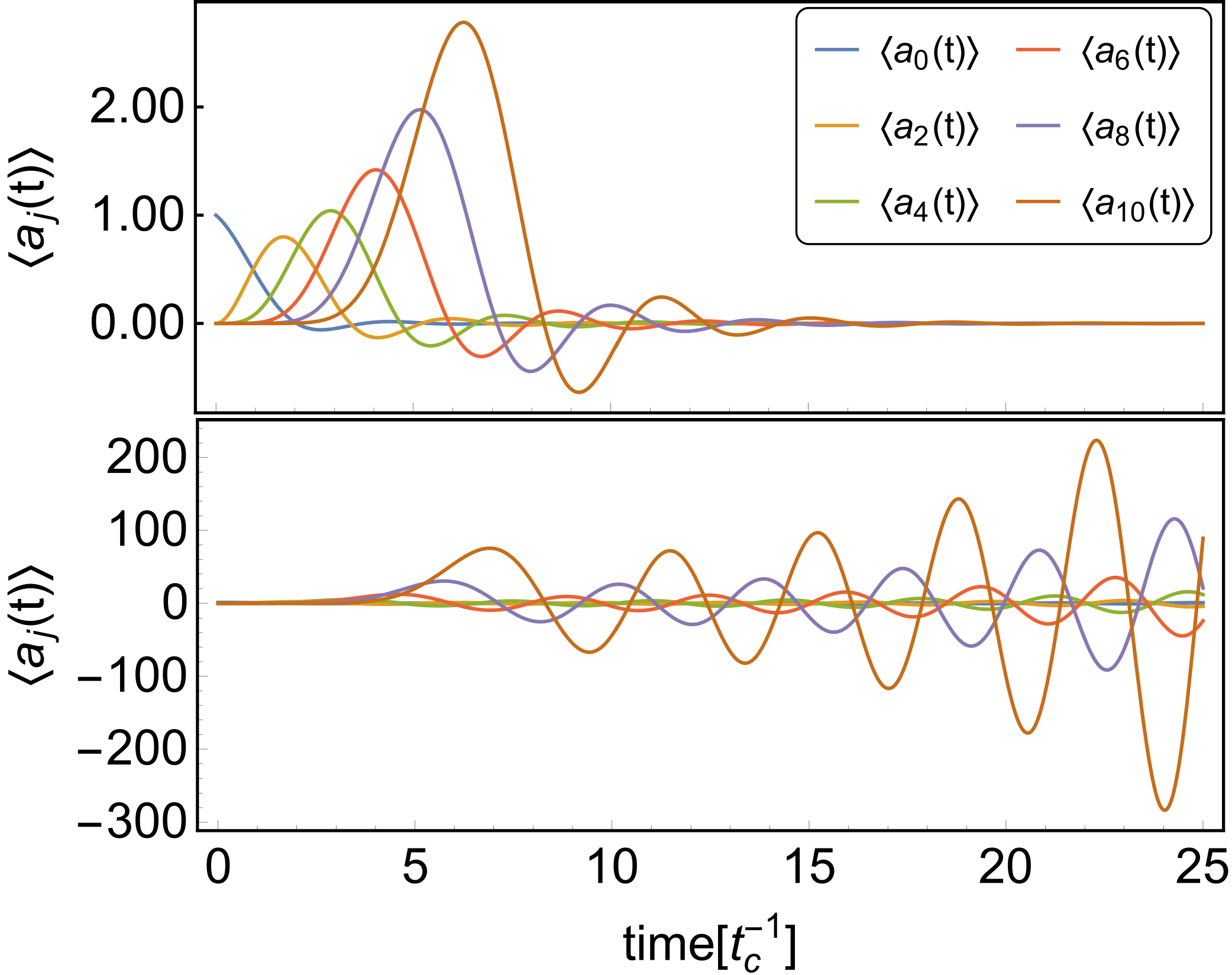}
    \caption{Regimes of amplification dynamics for the topological phase. The top plot shows the case with local dissipation dominating over collective pump ($\gamma/t_c=2$), where the excitation is amplified as it propagates, but rapidly damping after hopping to the next site. The bottom plot shows the regime dominated by collective pump ($\gamma/t_c=1$), where the excitation is delocalized between sites and constantly amplified over time. Parameters: $\phi=\pi/2$, $P/t_c=1.4$  and $\epsilon/t_c=0$.}
    \label{fig:Transient2}
\end{figure}

The other phase that can be considered is the one of topological amplification. Interestingly,  we find two different dynamical regimes within this phase, depending on how the process of topological amplification happens. The first possibility is a regime with dissipation dominating over pump ($\gamma > P$), where the initial signal is amplified as it moves through the array, but while it leaves a site, the signal gets rapidly damped, vanishing at long time, as shown in Fig.~\ref{fig:Transient2} (top). The second possibility corresponds to a regime with pump dominating over dissipation ($\gamma < P$), where the initial signal is amplified as it moves through the array, but the amplitude at each site keeps increasing over time, as shown in Fig.~\ref{fig:Transient2} (bottom). 
The main difference between these two regimes is the steady state solution, which in the first case has vanishing average at each site, while in the second one diverges.
This means that we can separate the phase of topological amplification in two dynamical regimes, where one is stable, while the other is unstable.
\begin{figure}[tb]
    \centering
    \includegraphics[width=1\columnwidth]{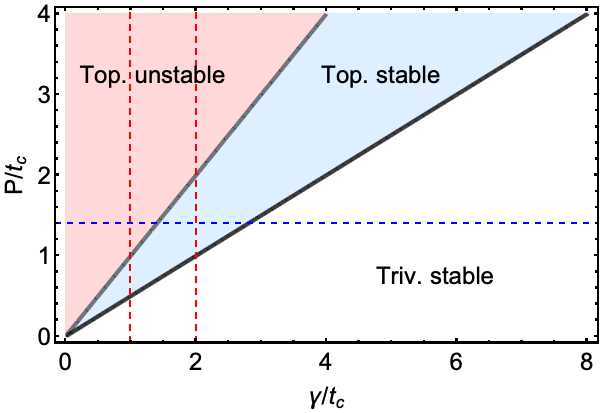}
    \caption{Stability diagram and its relation with the topological phase. The trivial phase is always stable (white region). In contrast, the topological phase is divided in two: a stable region (red) and an unstable one (blue). The crossing between dashed lines indicate the two points chosen in Fig.~\ref{fig:Transient2} to represent the different dynamical regimes of amplification. Parameters: $\phi=\pi/2$ and $\epsilon/t_c=0$.}
    \label{fig:Stability}
\end{figure}

This interesting behavior can be linked with the stability of the system, and demonstrates that, from the transient dynamics, it is possible to not just extract the presence of a topological amplification phase, but also to determine if it is stable or unstable and see how the instability develops over time. 
This is illustrated in Fig.~\ref{fig:Stability}, where we plot, as a function of the dissipative parameters $\gamma/t_c$ and $P/t_c$, the different possibilities found for this model.
The white region is characterized by a stable trivial phase, where amplification is absent and fluctuations are rapidly suppressed due to damping (see Fig.~\ref{fig:Transient1}). 
In contrast, coloured regions characterize the presence of topological amplification, which can be stable (blue) or unstable (red). In the unstable (topological) case, random fluctuations will be indefinitely amplified, as shown in Fig.~\ref{fig:Transient2} (bottom).
However, in the stable (topological) region, fluctuations are damped while signals are amplified.
The stability diagram has been obtained from the sign of the imaginary part of the eigenvalues of $\mathcal{D}$ for a finite system, and we have checked that it is independent of the system size, in agreement with Ref.~\cite{Ramos2021}. Hence, it links the different dynamical regimes obtained in the transient, with the steady state properties.

Finally, an interesting estimate from the transient dynamics is the propagation speed of the excitation or the average time that is required to amplify a signal. This is important to characterize amplifiers, because in addition to the steady state solution, which indicates the gain produced by the amplifier, it is also important to determine the time it takes that signal to be amplified. Obviously, one is interested in amplifiers that are fast.
To estimate the time it takes the excitation to travel from one site to the next, $\tau_{\text{amp}}$, we can estimate the distance between the zeroes of two consecutive Bessel functions. 
In the long time regime, it can be characterized from their average phase difference, resulting in $\tau_\text{amp} \sim\pi/|4 \sqrt{\alpha}|$, being $\alpha=\tilde{t}_{+}\tilde{t}_{-}=t_{c}^{2}-P^{2}/16+it_{c}P\cos\left(\phi\right)/2$, which perfectly agrees with the numerical results (notice that Fig.~\ref{fig:Transient2} is shown for even sites of the array only). 
This means that the amplifying time is thus inversely proportional to $|\tilde{t}_{+}\tilde{t}_{-}|$
, and thus, its value can be tuned to control the propagation velocity.
\subsubsection{Connecting Green's function with input-output theory~\label{subsec:amplification}}
So far, we have considered situations where the only driving of the systems comes either from an initial state or through incoherent pumping terms. Another relevant situation, especially for the purpose of amplification, is that in which there is an additional input field in some of the cavities. For those cases, it is convenient to adopt the input-output formalism, in which one writes a Langevin equation of the form~\cite{Ramos2021}:
\begin{equation}
    \dot{a}_j=-i \sum_{l}\mathcal{D}_{j,l}a_l+b_{\text{in},j}(t),\label{eq:EOM-fieldOp}
\end{equation}
to calculate the evolution of the system operators $a_j$ in the presence of a driving field $b_{\text{in},j}(t)$ at the $j$-site. $\mathcal{D}_{j,l}$ is the dynamical non-Hermitian matrix in Eq.~\eqref{eq:GF-EOM1}. If we are interested in the steady state properties, it is possible to apply a Fourier transform and solve for $a_j(\omega)$~\footnote{In contrast with~\cite{Ramos2021}, we consider the Green's function $G(\omega)$, which is related with their propagator by $G(\omega)=i Q(\omega)$}:
\begin{equation}
   a_j(\omega)=i \sum_{l}G_{j,l}(\omega)b_{\text{in},j}(\omega)\,,
\end{equation}
where we have used that the Green's function is the resolvent for the homogeneous part of the equation of motion. Thus, we can use the Green's functions obtained in Section~\ref{sec:decimation}, to find analytical expressions for the relation between the input and output fields.

\begin{figure} [tb]
    \centering
    \includegraphics[width=1\columnwidth]{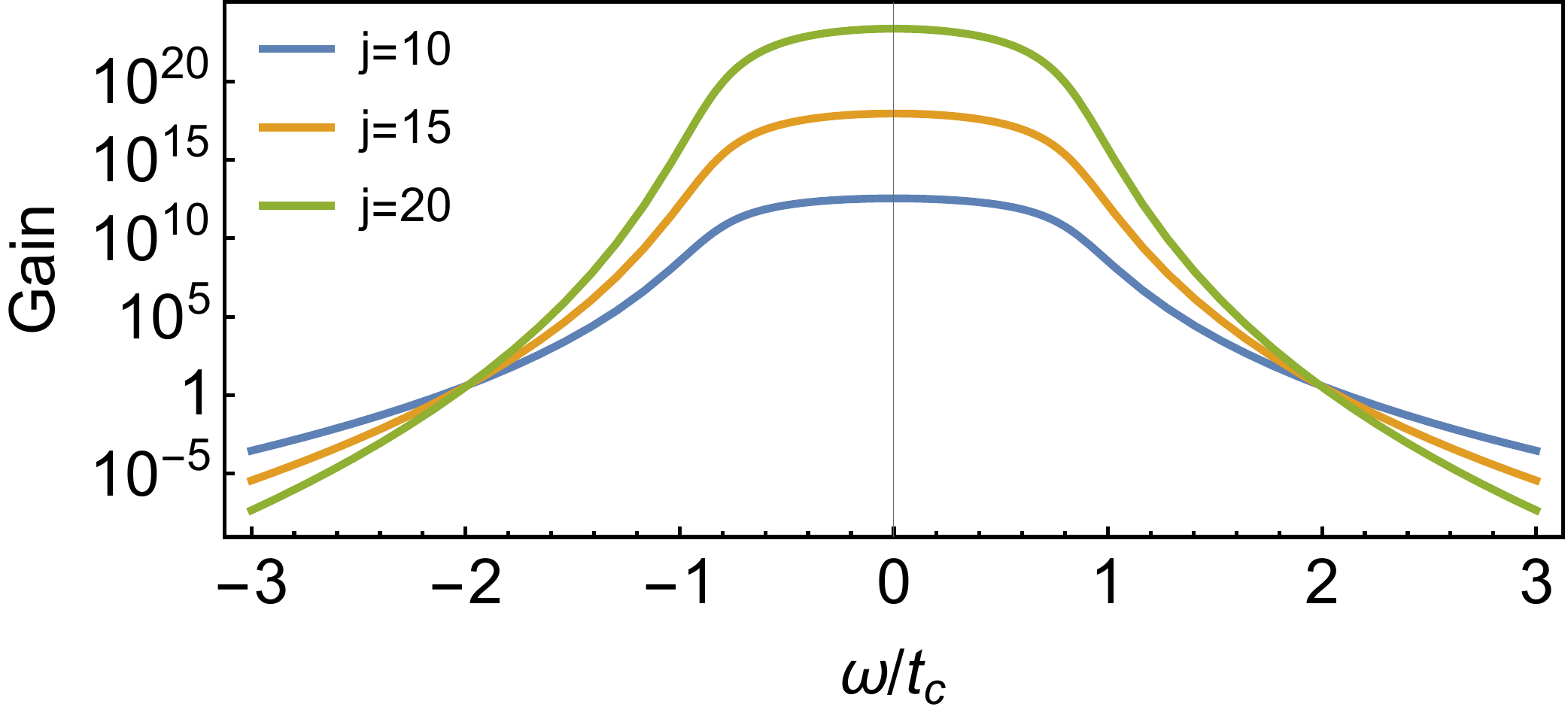}
    \includegraphics[width=1\columnwidth]{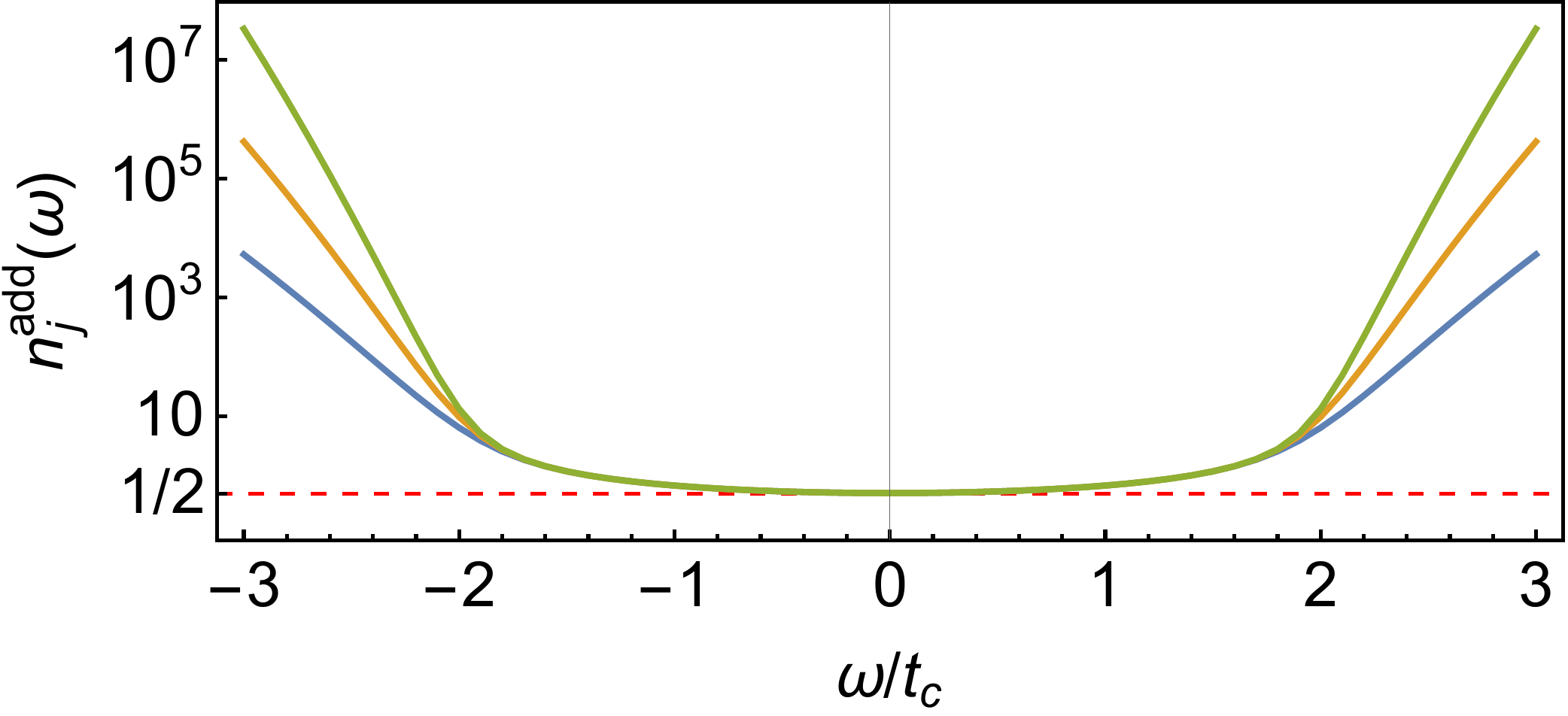}
    \caption{(Top) Gain vs $\omega$ at different sites. The input port is at the edge and as the signal propagates, gets exponentially amplified with the distance for a finite range of frequencies. (Bottom) Normalized added noise during the amplification process. The frequencies within the topological phase show a noise-to-signal ratio nearly at the quantum limit $n^{\text{add}}_{j}(\omega)=1/2$ (dashed line). Parameters: $\epsilon/t_c=0$, $\phi=\pi/2$, $\gamma/t_c=4$ and $P/t_c=3.6$.}
    \label{fig:Gain}
\end{figure}

For example, let us consider, that the input is a coherent state inserted at port $j=0$ (i.e., at the boundary of the chain). In that case, we can calculate the gain, $\mathcal{G}_j(\omega)$, at site $j>0$ as follows~\cite{Ramos2021}:
\begin{equation}
\mathcal{G}_{j}(\omega)=\gamma^2 \left| G_{j,0}(\omega) \right|^2=\gamma^2 |G_{0,0}(\omega)|^2 e^{2j\Re \xi(\omega)}\,,
\end{equation}
where we have used Eq.~\eqref{eq:simpleG} and $\xi(\omega)=\log \left[ G_{0,0}(\omega) \tilde{t}_{-} \right]$. Fig.~\ref{fig:Gain} (top) shows the gain as a function of $\omega$ for different lattice sites. As the plot is in logarithmic scale, we can confirm that the amplification process happens exponentially with the number of sites. Furthermore, there is a central plateau where the amplification is quite homogeneous for a wide range of frequencies.

Another interesting property, which can be analytically calculated, is related to how the photonic lattice changes the noise-to-signal ratio. The normalized added noise is defined as $n_j^{\text{add}}(\omega)=n_j^{\text{amp}}(\omega)/\mathcal{G}_j(\omega)$~\cite{caves81a},
with $n_j^{\text{amp}}(\omega)$ the added noise by the amplification process~\cite{Ramos2021} given by~\footnote{Notice that in Ref.~\cite{Ramos2021} the definition of added noise does not include the $1/2$ factor as it corresponds to a different convention for the vacuum noise}:
\begin{equation}
n^{\text{amp}}_j(\omega)=\frac{\gamma}{2}\sum_{l,l^\prime}G_{j,l}^*(\omega) G_{j,l^\prime}(\omega) P_{l,l^\prime}\,.\label{eq:AddedNoise}
\end{equation}
We plot $n^{\text{add}}_j(\omega)$ at various lattice sites in Fig.~\ref{fig:Gain} (bottom). This shows that the topological phase is an amplifier with an excellent noise-to-signal ratio, close to the quantum limit of $n^{\text{add}}_j(\omega)=1/2$.
To understand in simple terms why the phase of topological amplification is near the quantum limit, we can check the asymptotic behavior of $n^{\text{add}}_j(\omega)$ in Eq.~\eqref{eq:AddedNoise}. 
In the topological phase and for an input port at the edge, the sum is dominated by the Green's function with the largest number of sites contributing to amplification, $G_{j,0}(\omega)$ (contributions with $G_{j,l>j}$ are exponentially suppressed due to the directionality of the phase). 
In that case, we can approximate to lowest order $n^{\text{amp}}_j(\omega)\simeq \gamma P \left|G_{j,0}(\omega)\right|^2/2$. Therefore, if the system is tuned close to the center of the topological phase $\gamma \simeq P$ and is large enough, we find that $n^{\text{add}}_j(\omega)=n^{\text{amp}}_j(\omega)/\mathcal{G}_j(\omega)\to 1/2$ for all frequencies within the topologically non-trivial region. This result only relies on the system being a directional amplifier with exponential gain with the number of sites. Therefore, we can conclude that topological amplifiers will always operate near the quantum limit of the noise-to-signal ratio.

\section{Conclusions and outlook~\label{sec:conclu}}

To sump up, we introduce a new method to obtain the dynamical and steady-state properties of open quantum systems based on the calculation of lattice Green's functions using decimation techniques. Compared to other methods, our approach allows one to find compact analytical expressions in certain limits, which can bring additional understanding of the emergent phenomena. We have illustrated the power and versatility of the method with two examples of driven-dissipative bosonic chains, including one with a topological amplifier phase. Thanks to the analytical understanding brought by this method, we have been able to extract relevant observables, such as the signal-to-noise ratio, or magnitudes like coherence propagation length or the topological amplifying time. Besides, we have identified to different transient dynamical regime in the topological amplifying phase of the Hatano-Nelson model, which could not be extracted by the conventional steady-state analysis done in the literature.

An interesting outlook of this work would be to apply these tools to more complex scenarios, such as fermionic models or arrays of parametric amplifiers~\cite{Esposito2021PerspectiveAmplifiers,Sivak2020JosephsonAmplifier,White2015TravelingMatching,Macklin2015AAmplifier}.

\begin{acknowledgements}
We acknowledge support from CSIC Research Platform on Quantum Technologies PTI-001 and from Spanish project PGC2018-094792-B-100(MCIU/AEI/FEDER, EU), and from the Proyecto Sin\'ergico CAM 2020 Y2020/TCS-6545 (NanoQuCo-CM).
\end{acknowledgements}


~\newpage

\begin{widetext}
\appendix
\section{Derivation of the surface Green's function equation in method 2\label{ap:A}}
We start from the standard Dyson's equation for the Green's function, which separates the effective Hamiltonian into its perturbed and unperturbed parts: $\mathcal{D}=\mathcal{D}_0+\mathcal{V}$, where $\mathcal{V}$ describes the dissipative hopping between the last site of the array and its nearest neighbor, and $\mathcal{D}_0$ contains the local energy and dissipation, as well as all the other hopping terms.
With this separation, the Dyson's equation can be written as:
\begin{equation}
    G(\omega)=g(\omega)+g(\omega) \mathcal{V} G(\omega)\label{eq:decimation1},
\end{equation}
where we have defined the matrix of the unperturbed Green's function $g(\omega)=(\omega-\mathcal{D}_0)^{-1}$:
\begin{equation}
    g\left(\omega\right)=\left(\begin{array}{cccc}
g_{0,0}\left(\omega\right) & 0 & 0 & \ldots\\
0 & g_{1,1}\left(\omega\right) & g_{1,2}\left(\omega\right) & \ldots\\
0 & g_{2,1}\left(\omega\right) & g_{2,2}\left(\omega\right) & \ldots\\
\vdots & \vdots & \vdots & \ddots
\end{array}\right)\,,
\end{equation}
and the perturbation matrix:
\begin{equation}
    \mathcal{V}=\left(\begin{array}{cccc}
0 & \mathcal{V}_{+} & 0 & \ldots\\
\mathcal{V}_{-} & 0 & 0 & \ldots\\
0 & 0 & 0 & \ldots\\
\vdots & \vdots & \vdots & \ddots
\end{array}\right).
\end{equation}
If we now consider the matrix elements in Eq.~\eqref{eq:decimation1} and focus on the equation for the surface Green's function $G_{0,0}(\omega)$, we find:
\begin{equation}
    G_{0,0}\left(\omega\right)=g_{0,0}\left(\omega\right)+g_{0,0}\left(\omega\right)\mathcal{V}_{+}G_{1,0}\left(\omega\right)\,,\label{eq:decimation2}
\end{equation}
which is coupled to the equation for $G_{1,0}(\omega)$:
\begin{equation}
    G_{1,0}\left(\omega\right)=g_{1,1}\left(\omega\right)\mathcal{V}_{-}G_{0,0}\left(\omega\right)\,.
\end{equation}

Within this method, finding the solution is straightforward in the semi-infinite limit, by imposing that $g_{j+1,l+1}(\omega)=G_{j,l}(\omega)$. 
This makes Eq.~\eqref{eq:decimation2} transform into the non-linear equation:
\begin{equation}
    G_{0,0}\left(\omega\right)=g_{0,0}\left(\omega\right)+g_{0,0}\left(\omega\right)\mathcal{V}_{+}G_{0,0}\left(\omega\right)\mathcal{V}_{-}G_{0,0}\left(\omega\right).\label{eq:decimation3}
\end{equation}

Once Eq.~\eqref{eq:decimation3} is solved, it is possible to express an arbitrary Green's function in terms of $G_{0,0}(\omega)$ and the dissipative hopping $\mathcal{V}_{\pm}$.
For this, just notice that the following equations are also obtained from Eq.~\eqref{eq:decimation1}:
\begin{equation}
    G_{j,0}(\omega)=G_{j-1,0}(\omega)\mathcal{V}_{-}G_{0,0}(\omega)\to G_{j,0}\left(\omega\right)=\left[\mathcal{V}_{-}G_{0,0}\left(\omega\right)\right]^{j}G_{0,0}\left(\omega\right)\,.
\end{equation}

Similarly, for the other matrix elements one finds:
\begin{equation}
    G_{0,j}\left(\omega\right)=\left[G_{0,0}\left(\omega\right)\mathcal{V}_{+}\right]^{j}G_{0,0}\left(\omega\right)\,.
\end{equation}

Evaluating all the other matrix elements, one can deduce the form written in Eq.~\eqref{eq:Sol-G}:
\begin{equation}
    G_{j,l} = \left(G_{0,0}\mathcal{V}_{\text{sgn}\left(l-j\right)}\right)^{\left|j-l\right|}G_{0,0}+\sum_{a=0}^{\min\left\{j,l\right\}-1}\left(G_{0,0}\mathcal{V}_{-}\right)^{j-a}\left(G_{0,0}\mathcal{V}_{+}\right)^{l-a}G_{0,0}.
\end{equation}

\section{Analytical calculation of the Green's function for a 1D array\label{ap:B}}

In the 1D lossless case, the 2-point Green's function for the case of PBC can be analytically calculated. This requires to calculate the integral:
\begin{equation}
    G_{j,l}\left(\omega\right)=\int_{0}^{2\pi}\frac{dk}{2\pi}\frac{e^{ik\left(j-l\right)}}{\omega-2t_c\cos\left(k\right)}\,.
\end{equation}

The exact result is easily obtained using contour integral methods, which only require to define $z=e^{ik}$.
This substitution leads to the following integral form:
\begin{equation}
    G_{j,l}\left(\omega\right)=\frac{1}{2\pi i}\oint\frac{z^{j-l}dz}{z\omega-t_cz^{2}-t_c}\,,
\end{equation}

The denominator has two poles at $z_{\pm}=\frac{\omega\pm\sqrt{\omega^{2}-4t_c^{2}}}{2t_c}$, which can be used to calculate the integral, noticing that only $z_{-}$ is within the unit circle.
The result separates two regions:
\begin{equation}
    G_{j,l}\left(\omega\right)=\begin{cases}
\frac{\text{sign}\left(\omega\right)}{\sqrt{\omega^{2}-4t_c^{2}}}\left[\frac{\omega}{2t_c}-\text{sign}\left(\omega\right)\sqrt{\left(\frac{\omega}{2t_c}\right)^{2}-1}\right]^{\left|j-l\right|} & \text{for }\left|\omega\right|>2t_c\,,\\
\frac{\mp i}{\sqrt{\omega^{2}-4t_c^{2}}}\left[\frac{\omega}{2t_c}\pm\frac{1}{i}\sqrt{1-\left(\frac{\omega}{2t_c}\right)^{2}}\right]^{\left|j-l\right|} & \text{for }\left|\omega\right|<2t_c\,,
\end{cases}
\end{equation}
where the $\pm$ corresponds to the retarded/advanced Green's function, respectively.
This expression describes the exact Green's function for the Hermitian 1D tight-binding model. To write it as in Eq.~\eqref{eq:Gijbulk} we just need to exponentiate it and take its logarithm. This allows to encode the spatial dependence in the exponential.
\section{Derivation of the inverse Laplace transform\label{ap:C}}
The solution for the surface Green's function is:
\begin{equation}
G_{0,0}\left(s\right)=\frac{1-\sqrt{1-4g_{0,0}\left(s\right)^{2}\tilde{t}_{+}\tilde{t}_{-}}}{2g_{0,0}\left(s\right)\tilde{t}_{+}\tilde{t}_{-}}\,.
\end{equation}

Now, we define $2g_{0,0}\left(s\right)=y\left(s\right)$ and $z\left(y\right)=\sqrt{1-y^{2}\alpha}$,
being $\alpha=\tilde{t}_{+}\tilde{t}_{-}$. With this we can write
the required expression as:
\begin{align}
G_{0,0}\left(s\right)^{x+1} & =\left(y\alpha\right)^{-x-1}\left(1-z\right)^{x+1}\nonumber \\
 & =\left(y\alpha\right)^{-x-1}\sum_{n=0}^{\infty}\left(\begin{array}{c}
x+1\\
n
\end{array}\right)\left(-1\right)^{n}z^{n}\nonumber \\
 & =\left(y\alpha\right)^{-x-1}\sum_{n=0}^{\infty}\left(\begin{array}{c}
x+1\\
n
\end{array}\right)\left(-1\right)^{n}\sum_{p=0}^{\infty}\left(\begin{array}{c}
n/2\\
p
\end{array}\right)\left(-1\right)^{p}\alpha^{p}y^{2p}\nonumber \\
 & =\sum_{n=0}^{\infty}\left(\begin{array}{c}
x+1\\
n
\end{array}\right)\left(-1\right)^{n}\sum_{p=0}^{\infty}\left(\begin{array}{c}
n/2\\
p
\end{array}\right)\left(-1\right)^{p}\alpha^{p-x-1}y^{2p-x-1}\,.
\end{align}

Now, we can perform the inverse Laplace transform of each term:
\begin{equation}
\mathcal{L}^{-1}\left\{ y\left(s\right)^{2p-x-1}\right\} \left(t\right)=-i2^{2p-x-1}\frac{e^{-it\tilde{\epsilon}}\left(-it\right)^{2p-x-2}}{\Gamma\left(2p-x-1\right)}\,,
\end{equation}
and use this result to write the inverse Laplace transform of the
powers of the Green's function in time-domain:
\begin{align}
G_{0,0}\left(t\right)^{x+1} & =-ie^{-it\tilde{\epsilon}}\sum_{n=0}^{\infty}\left(\begin{array}{c}
x+1\\
n
\end{array}\right)\left(-1\right)^{n}\sum_{p=0}^{\infty}\left(\begin{array}{c}
n/2\\
p
\end{array}\right)\left(-1\right)^{p}\frac{\alpha^{p-x-1}2^{2p-x-1}\left(-it\right)^{2p-x-2}}{\Gamma\left(2p-x-1\right)}\nonumber \\
 & =e^{-it\tilde{\epsilon}}\left(-i\right)^{x+1}\left(x+1\right)\frac{\mathcal{J}_{x+1}\left(2t\sqrt{\alpha}\right)}{t\alpha^{\frac{x+1}{2}}}\,,
\end{align}
where $\mathcal{J}_{r}\left(x\right)$ is the r-th Bessel function
of the first kind.

Finally, we can write the time-evolution of the average of the field
operator at site $x$ as:
\begin{equation}
\langle a_{x}\left(t\right)\rangle=i\langle a_{0}\rangle e^{-it\tilde{\epsilon}}\left(-i\right)^{x+1}\left(x+1\right)\left(\frac{\tilde{t}_{-}}{\tilde{t}_{+}}\right)^{x/2}\frac{\mathcal{J}_{x+1}\left(2t\sqrt{\tilde{t}_{+}\tilde{t}_{-}}\right)}{t\sqrt{\tilde{t}_{+}\tilde{t}_{-}}}\,.
\end{equation}

\end{widetext}


\bibliographystyle{apsrev4-2}
\bibliography{references}

\end{document}